\documentclass[11pt]{article}
\usepackage{float}
\usepackage{amsmath, mathrsfs, amsthm}
\usepackage{amssymb}
\usepackage{booktabs}
\usepackage{mathtools}
\usepackage{soul}
\usepackage{hyperref}
\usepackage{dsfont}

\usepackage{bm}
\usepackage{natbib}
\usepackage[capitalise,noabbrev]{cleveref}

\usepackage[plain,noend]{algorithm2e}
\usepackage{subcaption}

\usepackage{graphicx}

\usepackage{mathtools}

\allowdisplaybreaks

\usepackage{color}

\newtheorem{theorem}{Theorem} 
\newtheorem{lemma}{Lemma}

\newcommand{\be}{\begin{equation*}\begin{aligned} }
\newcommand{\ee}{\end{aligned}\end{equation*} }

\newcommand{\bel}{\begin{equation}\begin{aligned} }
\newcommand{\eel}{\end{aligned}\end{equation} }
\renewcommand{\t}[1]{\text{#1} }
\renewcommand{\Pr}{p}

 \DeclareMathOperator{\No}{N}

\DeclareMathOperator{\Dir}{Dir}
\DeclareMathOperator{\Wishart}{Wishart}
\DeclareMathOperator{\Ga}{Gamma}

\let\oldproofname=\proofname
\renewcommand{\proofname}{\rm\bf{\oldproofname}}

\begin{document}

\title{Consistent Model-based Clustering\\
using the Quasi-Bernoulli Stick-Breaking Process}

\author{Cheng Zeng
	\thanks{Department of Statistics, University of Florida, U.S.A. {czeng1@ufl.edu}}\quad\quad
	Jeffrey W. Miller
	\thanks{Department of Biostatistics , Harvard University, U.S.A. {jwmiller@hsph.harvard.edu}}\quad\quad
	Leo L. Duan
	\thanks{Department of Statistics, University of Florida, U.S.A. {li.duan@ufl.edu}}}

\maketitle

\begin{abstract}%
In mixture modeling and clustering applications, the number of components and clusters is often not known. A stick-breaking mixture model, such as the Dirichlet process mixture model, is an appealing construction that assumes infinitely many components, while shrinking the weights of most of the unused components to near zero. However, it is well-known that this shrinkage is inadequate: even when the component distribution is correctly specified, spurious weights appear and give an inconsistent estimate of the number of clusters. In this article, we propose a simple solution: when breaking each mixture weight stick into two pieces, the length of the second piece is multiplied by a quasi-Bernoulli random variable, taking value one or a small constant close to zero. This effectively creates a soft-truncation and further shrinks the unused weights. Asymptotically, we show that as long as this small constant  diminishes to zero at a rate faster than $o(1/n^2)$, with $n$ the sample size, the posterior distribution will converge to the true number of clusters. In comparison, we rigorously explore Dirichlet process mixture models using a concentration parameter that is either constant or rapidly diminishes to zero --- both of which lead to inconsistency for the number of clusters. Our proposed model is easy to implement, requiring only a small modification of a standard Gibbs sampler for mixture models. In simulations and a data application of clustering brain networks, our proposed method recovers the ground-truth number of clusters, and leads to a small number of clusters.
\end{abstract}
        

% \usepackage{bm}
% \usepackage{natbib}

% \usepackage{enumerate}
% \usepackage{verbatim}

% %\openup 1em

% %macro for commenting
% \usepackage{color}

% \newcommand{\leo}[1]{{\color{red}{#1}}}
% \newcommand{\cheng}[1]{{\color{blue}{#1}}}

% % \newcommand{\Xbeta}{ X_i \theta}
% \newcommand{\xbeta}{ x_i \beta} \newcommand{\xtheta}{ x_i \theta}
% % \newcommand{\xbetaij}{ x_{ij}^T \theta}
% \newcommand{\sgamma}{s_{ij}^T\gamma_i} \newcommand{\core}{constraint relaxation}

% \usepackage{rotating} \usepackage{graphicx}

% \usepackage{float}
% %\usepackage{bbm}
% %\usepackage{dsfont}

% \usepackage{amsmath, amssymb, amsthm}
% \usepackage{mathrsfs}
% \usepackage{subcaption}
% \usepackage[normalem]{ulem}
% %\usepackage{nicefrac}

% \usepackage{tikz}
% \usepackage{tikz-3dplot}

% \usepackage{hyperref}

\section{Introduction}

Mixture models are frequently used to analyze data with unknown group/cluster structure. They give a generative view on the data $y = y_{1:n} = (y_1,\ldots,y_n)$, and provide uncertainty quantification on the cluster assignments. To review the main idea, suppose
\[
y_i \stackrel{iid}\sim\sum_{k=1}^{K} w_k \mathcal{F}(\theta_k)
\]
for $i=1,\dots,n$, where $\mathcal{F}(\theta_k)$ is a distribution parameterized by $\theta_k$, and $w_1,\ldots,w_K\geq 0$ such that $\sum_{k=1}^K w_k = 1$. Using the Bayesian framework {that posits a prior distribution for the weights $w= w_{1:K} = (w_1,\ldots,w_K)$ and the parameters $\theta = \theta_{1:K} = (\theta_1, \ldots, \theta_K)$}, one can infer the posterior distribution of the weights $w $ and the parameters $\theta $, as well as the assignments of data points to mixture components, which yields a clustering of the data \citep{fraley2002model}.

In practice, {in addition to $w_k$ and $\theta_k$}, we usually do not know the number of clusters either. Stick-breaking models provide an appealing solution in which the number of mixture components $K$ is  infinite, and the number of clusters  in the data (that is, the number of components that the data are assigned to) can be any finite number. A general form of stick-breaking model for the mixture weights $w$ is
\bel\label{eq:canonical_sb}
w_1 = v_1 \text{ and } w_k = v_k\prod_{l=1}^{k-1} (1-v_l) \text{ for $k=2,3,\ldots,$}
\eel
where $v_1,v_2,\ldots$ are drawn from some prior distribution.
The interpretation is that starting from a stick of length $1$, for each $k$ we break off a proportion $v_k\in [0,1]$ from the remaining stick, and use it as the weight $w_k$.

Many priors have been proposed for the distribution of $v_k$. Perhaps the most widely used one is $v_k\sim \text{Beta}(1,\alpha)$, {with which the probability distribution $\sum_{k=1}^\infty w_k \delta_{\theta_k}(\cdot)$ yields an equivalent realization of the} Dirichlet process with concentration parameter $\alpha$ \citep{sethuraman1994constructive}, {where $\delta_x(\cdot)$ represents the Dirac measure which satisfies $\delta_x(A)=\mathds{1}(x\in A)$ for any measurable set $A$.}  More generally, when $v_k\sim \text{Beta}(1-d,\alpha + k d)$ {with $0<d<1$ and $\alpha>-d$}, one obtains the Pitman--Yor process {with discount parameter $d$ and strength parameter $\alpha$} \citep{pitman1997two, ishwaran2001gibbs}. 

%However, it has been discovered that such a shrinkage is not optimal: as the data sample size increases, more and more spurious small clusters will appear.
% A long-held belief is that this is a model misspecification problem in the component distribution, which has motivated a large class of models focused on calibrating $\mathcal F$: for example, \cite{kosmidis2014model} uses copulas (modeling cumulative distribution function) to replace standard parametric $\mathcal F$; \cite{miller2019robust} replaces $\mathcal F$ with a tempered likelihood, giving some tolerance to model misspecification; \cite{xie2020bayesian} regularizes the minimum distance between any two $\theta_k$'s away from zero, preventing the creation of new cluster with trivial difference from existing ones. Despite some empirical improvements, it is unknown whether these methods can lead to to an optimal estimate on the cluster number [although consistency can be achieved on the density estimation \citep{ghosal1999posterior}, it does not apply on the number of clusters].

When the true data-generating distribution is a finite mixture from the assumed family, these models tend to shrink most of $w_k$'s close to zero, leading to a small number of clusters in the posterior.
However, \cite{miller2013simple,miller2014inconsistency} showed a striking result: neither the Dirichlet process prior nor the Pitman--Yor process prior {allows the posterior distribution on the number of components $K$} to converge to the true number of clusters as $n$ increases, even when the family of component distributions $\mathcal{F}$ is correctly specified.

To address this issue, we develop a prior on the $w_k$'s that yields stronger shrinkage while remaining easy to use. Specifically, we modify the canonical stick-breaking construction as follows: at each break, we multiply the remaining proportion $(1-v_k)$ by a discrete random variable that takes value either $1$ or $\epsilon$. When the latter happens, the tail probability $\sum_{k=L+1}^{\infty} w_k$ is strictly bounded by $\epsilon$. We show that if $\epsilon$ is chosen in a sample-size-dependent way such that  $\epsilon(n) = o(1/n^{2+r})$ (with $r$ a non-negative integer that depends on $\alpha$), then one can obtain posterior consistency for the number of clusters. In the special case of $\epsilon = 0$, this model reduces to a finite mixture model with a prior on the number of components \citep{miller2018mixture}, which also yields posterior consistency for the number of clusters. Meanwhile, in the special case of $\epsilon = 1$ and $v_k \sim \t{Beta}(1,\alpha)$, this model reduces to a Dirichlet process mixture.

% We are inspired by the common heuristic of merging small clusters together after model fitting, as routinely used in the statistical applications \citep{stephenson2020robust}. To achieve a similar effect in a model-based framework, we need a generative mechanism to produce a strict bound on the tail probability of the form $\sum_{k=L+1}^{\infty} w_k < \epsilon$ with $\epsilon$ small.

Therefore,  using $\epsilon \in (0,1)$ effectively interpolates between these two extremes. Compared to $\epsilon = 0$, using $\epsilon\in(0,1)$ avoids having a singularity at $w_k = 0$. This relaxes the parameter space in a way that allows the Markov chain Monte Carlo (MCMC) sampler to more efficiently add and remove clusters, rather than employing an explicit discrete search over $K$. Further, it makes the technique compatible with more complex stick-breaking models, such as the ones involving kernels \citep{dunson2008kernel}, geospatial processes \citep{rodriguez2010latent}, and external predictors \citep{ren2011logistic}. On the other hand, compared to $\epsilon = 1$, it allows the model to behave effectively like a finite mixture with a prior on the number of components, leading to superior control on the number of clusters.
%The Dirichlet process inconsistency results of \cite{miller2013simple,miller2014inconsistency} are for the case of fixed $\alpha$, and it is plausible that a Dirichlet process mixture might yield consistency for the number of clusters if one chooses
%$\alpha$ in a sample-size-dependent way such that $\alpha \to 0$ at an appropriate rate as $n\to\infty$.  
%However, even if this is possible, it would result in a prior on weights $w_k$ that very strongly favors low entropy configurations; 
%thus, for instance, having $K$ roughly equally sized weights is strongly disfavored when $\alpha$ is small.
%Alternatively, consistency may be possible by placing a prior on $\alpha$ that allows the posterior on $\alpha$ to concentrate near zero as $n$ grows, however, this would still have the same low entropy issue.
%Further, we show that if $\alpha$ goes to zero too fast, then the Dirichlet process mixture still exhibits inconsistency.
%Thus, it does not seem possible to fully resolve the issue by playing with $\alpha$ when using a Dirichlet process.
We illustrate these advantages in simulations and a data application in clustering brain networks,  using a mixture of low-rank probit models.
A software implementation {and the steps needed to replicate the results in this paper} are provided on \href{https://github.com/Zeng-Cheng/quasi-bernoulli-stick-breaking}{\color{blue} https://github.com/zeng-cheng/quasi-bernoulli-stick-breaking}.

\section{Quasi-Bernoulli Stick-Breaking Process}\label{sec:QB}

{In this section, we formally construct the quasi-Bernoulli stick-breaking process, and provide theoretical analysis of this process and the corresponding mixture model.}

\subsection{Prior Construction}
%The stick-breaking process is a generative process for the parameters $\theta^*_i\ (i=1,\dots, n)$, as from a discrete distribution taking value $\theta_k$ with probability $w_k$. Starting from a stick of length $1$, each time we break a proportion of $v_k\in (0,1)$ away from the remaining stick, and use its length as $w_k$.
In the general form of the stick-breaking construction (\cref{eq:canonical_sb}), if the proportion $v_L$ at step $L$ is very close to $1$, then $w_L$ will take almost all the remaining stick, resulting in $w_k\approx 0$ for $k\geq L+1$. Based on this intuition, we introduce the following stick-breaking process:
\begin{equation}
	\begin{aligned}
		\label{eq:qbsb}
		w_1 &= v_1, \quad w_k = v_k\prod_{l=1}^{k-1} (1-v_l), \quad k\geq 2, \\
		v_k &= 1- b_k\beta_k ,\\
		b_k &\sim \tilde{p}\delta_1(\cdot) + (1-\tilde{p}) \delta_\epsilon(\cdot),\\
		\beta_k &\sim  \text{Beta}(\alpha, 1).
	\end{aligned}
\end{equation}
%where each $\theta_k\sim \mathcal G$, a certain base measure. 
%\cheng{where the $\delta_x(\cdot)$ is the Dirac probability measure which satisfies $\delta_x(A)=\mathds{1}(x\in A)$ for any measurable set $A$.}
Each $b_k$ follows a discrete distribution such that $b_k = 1$ with probability $\tilde{p}\in(0,1)$, and $b_k = \epsilon$ with probability $1-\tilde{p}$, for some small $\epsilon\in(0,1)$. We refer to $b_k$ as a quasi-Bernoulli random variable, since it resembles the standard Bernoulli supported on  $\{0,1\}$. We refer to \cref{eq:qbsb} as the quasi-Bernoulli stick-breaking process (or simply the ``quasi-Bernoulli process'').
%, denoted by  $\theta_i^* \sim \text{QB}_{\epsilon,p}(\alpha,\mathcal{G})$.
With this prior on weights $w_1,w_2,\ldots$, we obtain an infinite mixture model by letting $y_i \stackrel{iid}\sim \sum_{k=1}^\infty w_k \mathcal{F}(\theta_k)$, where $\theta_k\stackrel{iid}\sim \mathcal{G}$ from some base measure $\mathcal{G}$; we refer to this as a quasi-Bernoulli mixture model.
{Moreover, this marginal representation of the mixture model can be equivalently represented in a conditional form with introducing the latent assignment variable $c_i$ for each data point $y_i$.}
%, and each component with   $n_k=\sum_{i=1}^n 1(c_i=k)$ and $n_k > 0$ is referred to as a ``cluster''.
Specifically, with $c_i=k$ representing the event that $y_i$ is drawn from the mixture component $k$,
\bel\label{eq:clusters}
c_1,\dots,c_n \mid {w} &  \stackrel{iid}\sim \text{Categorical}({w}) ,\\
\theta_1,\theta_2,\ldots & \stackrel{iid}\sim \mathcal{G} ,\\
Y_i \mid c_i,{\theta}  & \sim \mathcal{F}(\theta_{c_i})\text{ independently for } i=1,\dots,n.\\
\eel
{This conditional representation is useful if one is interested in using the model to perform model-based clustering.}

Note that if $\epsilon=1$, then we would have $v_k=1-\beta_k \sim \t{Beta}(1,\alpha)$, yielding the stick-breaking representation of the Dirichlet process $\text{DP}(\alpha,\mathcal{G})$; whereas if $\epsilon=0$, then we would have a random truncation on $(w_1,w_2,\ldots)$. Using $\epsilon\in (0,1)$ yields a soft truncation that provides advantages from both of these extremes.
%, leading to both posterior consistency and a support for an infinite mixture model.

%For later reference, we note here that by the change of variables formula, the density of $v_k$ is
%\bel\label{eq:p_v}
%f_V(v) := p \t{Beta}(1-v \mid \alpha,1) + (1-p)\frac{1}{\epsilon}\t{Beta}\Big(\frac{1-v}{\epsilon}\mid \alpha,1\Big)
%\eel
%for all $k$,
%and further, since $\prod_{l=1}^{k-1} (1-v_l) = 1 - \sum_{l=1}^{k-1} w_l$, we have
%\bel\label{eq:p_w}
%p(w_k\mid w_1,\ldots,w_{k-1}) = \frac{1}{1 - \sum_{l=1}^{k-1} w_l} f_V\Big(\frac{w_k}{1 - \sum_{l=1}^{k-1} w_l}\Big).
%\eel

Before we move into more technical discussions, we first illustrate an intuition for why the Dirichlet process with a fixed $\alpha$ tends to create small and spurious clusters, and how our proposed prior mitigates this. 
Consider the scenario where we have $n$ data points already assigned to $K$ clusters, and there are $m$ new data points to be assigned. {Assume that there are only $K$ clusters in the population, but we are
	modeling the data as drawn from a Dirichlet process mixture model.} Ideally we want to assign all of the $m$ new data into the existing $K$ clusters. The prior probability of adding one or more new clusters can be calculated using the predictive rule (as in the ``restaurant process'')  $\Pr(c_i \mid c_{1},\ldots, c_{i-1})$ {for $i\in\{n+1,\dots,n+m\}$} recursively for $m$ times. For the Dirichlet process with concentration parameter $\alpha$, this probability has a closed-form:
\be
\Pr\left(\sum_{l=1}^m \mathds{1}(c_{n+l}>K) > 0  \mid c_1,\ldots, c_n\right) = 1- \prod_{l=1}^m \left(\frac{n+l-1}{n+l-1+\alpha}\right ),
\ee
{which follows directly from the predictive distribution under the Dirichlet process \citep[Equation (2)]{blackwell1973ferguson}.}
Although the probability is small for $m=1$, it increases rapidly and becomes non-trivial as $m$ grows.  Note that the above event includes not only assigning all $m$ data points into a single new cluster, but also having them scattered into several new clusters; hence this is the union of all undesirable outcomes of having spurious clusters.

\begin{figure}
	\centering
	\includegraphics[width=.7\linewidth]{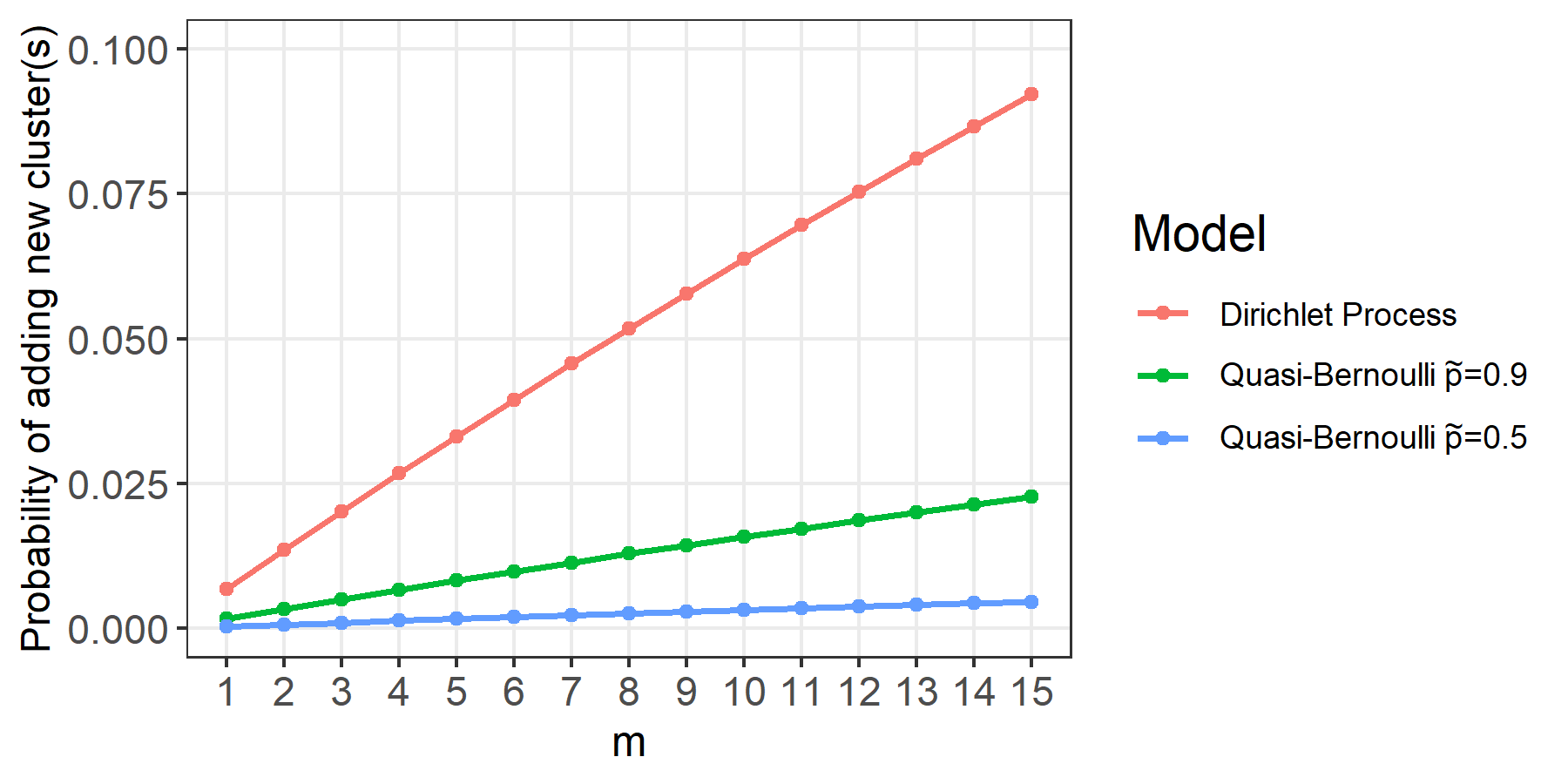}
	\caption{
		Under the prior, the Dirichlet process exhibits rapid growth in the probability of adding one or more new clusters for $m$ future data points ($n=100$), favoring the creation of additional clusters \textit{a priori}. Meanwhile, the quasi-Bernoulli process exhibits much slower growth of this probability. {For the quasi-Bernoulli process, we use $\epsilon=\epsilon(n,m)=1/(n+m)^{2.1}$ and $\alpha=1$ as suggested in our theory \cref{thm:consistent}. For the Dirichlet process, we use $\alpha=0.69$, which has the prior expected number of clusters close to the one under the quasi-Bernoulli process.}
		\label{fig:comp_w_dp} }
\end{figure}

For the quasi-Bernoulli process, although we do not have a simple closed-form expression for the above probability, we can numerically calculate it based on the partition probability function \cref{eq:eppf} introduced in the next section. In \cref{fig:comp_w_dp}, we examine the case when given two existing clusters with $n_1=n_2=50$, and assigning additional $m$ data points. It can be seen that
the prior probability of creating new cluster(s) quickly increases in the Dirichlet process,
%with $\alpha=0.69$, under which the Dirichlet process prior and the quasi-Bernoulli process prior have almost the same expectations on the number of clusters for the cases $n=100+m$
whereas the quasi-Bernoulli processes (with $\alpha=1$ 
% and $\epsilon = 1 / (100+m)^{2.1}$, 
and two values for $\tilde{p}$) substantially slow down the growth of this probability. {We also provide results with another rate of $\epsilon(n)$ in the appendix.}

We will carefully examine the posterior behaviors of the above models, including comparing with the Dirichlet process with $\alpha=\alpha(n)$ tending to zero as $n\to\infty$.

\subsection{Exchangeable Random Partitioning}\label{subsec:exch}

A large class of stick-breaking models enjoys the property of partition exchangeability
--- that is, the probability distribution of the induced partition only depends on the cluster sizes, and is invariant to any permutation of the cluster index \citep{pitman1995exchangeable}. 
Letting $i\in\{1,\dots,n\}$ be the data index, if there are  $t$ unique values in the cluster assignments  $c = (c_1,\dots,c_n)$, then we can form a corresponding partition $\mathcal{A} = \{A_1,\dots, A_t\}$ in the following way: (1) initialize $A_1 = \{1\}$ and $t=1$; (2) sequentially for $i=2,\dots,n$, if $c_i=c_j$ for any $j\le i-1$ and $j\in A_k$, then add $i$ to the same set $A_k$ containing $j$; otherwise create a new set $A_{t+1}=\{i\}$ and increment $t$ to $t+1$.

%It is not hard to see that the permutation of $[n]$ is equivalent to the permutation of $\{S_1,\ldots, S_t\}$. Therefore, we marginalize over all permutations to obtain the exchangeable partition probability function.
\begin{theorem}
	[Exchangeable Partition Probability Function]
	\label{thm:eppf}
	The probability mass function of the random partition $\mathcal{A} = \{A_1,\dots,A_t \}$ induced by $c$ in the quasi-Bernoulli stick-breaking process is
	\bel\label{eq:eppf}
	\Pr_{\epsilon,n}(\mathcal{A}) = \frac {\alpha^t\Gamma(\alpha)} {\Gamma(n+\alpha)} \biggl(\prod_{j=1}^t \Gamma(n_j+1)\biggr)
	\sum_{\sigma\in S_t}\prod_{j=1}^{t} \frac{\tilde{p} + (1-\tilde{p}) I_\epsilon(g_{j+1}(\sigma)+\alpha, n_{\sigma_j}+1)/\epsilon^\alpha}{g_j(\sigma)+\alpha(1 - \tilde{p})(1 - \epsilon^{g_j(\sigma)})}
	\eel
	where ${\sigma}= (\sigma_1,\dots,\sigma_t)$ is a permutation of $\{1,\dots,t\}$, $S_t$ is the set of all permutations of $\{1,\dots,t\}$, $n_j=|A_j|$, $g_j(\sigma) = \sum_{l=j}^t n_{\sigma_l}$, $\Gamma(x)=\int_0^\infty z^{x-1}e^{-z}\,dz$ and $I_\epsilon(q_{1},q_{2})$ is the cumulative distribution function of $\t{Beta}(q_{1},q_{2})$ evaluated at $\epsilon$.
\end{theorem}
%\begin{remark}
%        If $p=1$, the above function becomes the same as the one of the Chinese restaurant process.
%\end{remark}
For conciseness, we defer all the proofs to the appendix.
Using $\Pr_{\epsilon,n}(\mathcal{A})$, we can substantially simplify the model in \cref{eq:qbsb,eq:clusters} into an equivalent generative process:
\bel\label{eq:qbcluster}
\mathcal{A} & \sim \Pr_{\epsilon,n}(\mathcal{A}), \\
\theta_1,\ldots,\theta_t \mid \mathcal{A} & \stackrel{iid}\sim \mathcal{G} ,\\
y_i \mid \mathcal{A},\theta_1,\dots,\theta_t & \sim \mathcal{F}(\theta_j) \text{ for $i\in A_j$, $A_j\in\mathcal{A}$}.
\eel
We now use the above representation to study the asymptotics of the clustering when $n\to \infty$.

\subsection{Consistent Estimation of the Number of Clusters}\label{subsec:effe}

By definition, the number of clusters is $t=|\mathcal{A}|$.  We use $T$ to denote the associated random variable in the model. Let $\mathcal{H}_t(n)$ denote the set of all partitions of $\{1,\dots,n\}$ into $t$ disjoint sets. Using \cref{eq:qbcluster}, the marginal posterior of $T$ is
\begin{equation}\label{eq:postt}
	\begin{aligned}
		\Pr_{\epsilon} (T=t\mid{y_{1:n}}) = \frac {\sum_{\mathcal{A}\in \mathcal{H}_t(n)} p (y_{1:n}\mid \mathcal{A}) \Pr_{\epsilon,n} (\mathcal{A})} {\sum_{\mathcal{A}\in \cup_{t=1}^n \mathcal{H}_t(n)} p (y_{1:n} \mid \mathcal{A})\Pr_{\epsilon,n} (\mathcal{A}) } ,
	\end{aligned}
\end{equation}
where $p ({y_{1:n}} \mid \mathcal{A}) = \prod_{A\in\mathcal{A}} m(y_A)$, $y_A = (y_i:i\in A)$, and $m(y_A) = \int_{\Theta}\bigl( \prod_{i\in A} f_{\theta}(y_i) \bigr)\, d\mathcal{G}(\theta)$.
Here, $f_\theta$ denotes the density of the component distribution $\mathcal{F}(\theta)$.

Suppose the data are in fact generated from $k_0$ mixture components, with $k_0$ a fixed and finite number. We establish general conditions under which $\Pr_{\epsilon} (T=k_0\mid{y_{1:n}})\to 1$ as $n\to\infty$. Our proof involves two main parts: (i) we establish that this consistency property holds for the finite-dimensional model obtained by setting $\epsilon=0$; and (ii) we bound the total variation distance between the prior distributions $\Pr_{\epsilon,n}(\mathcal A)$ and $\Pr_{0,n}(\mathcal A)$.  We then show that this implies posterior consistency.

Consider the case when  $\epsilon=0$, that is, when $b_k$ in \cref{eq:qbsb} is a Bernoulli random variable with success rate $\tilde{p}$. {In this case, we refer to \cref{eq:qbsb} with  $\epsilon=0$ as the Bernoulli stick-breaking process.} Let $K = \min\{k : b_k = 0\}$. When this occurs, we have $w_{K+1}=w_{K+2}=\dots=0$, and therefore, $w$ is effectively $K$-dimensional.  Observe that $K$ follows a geometric distribution: $\pi_K(k) = \tilde{p}^{k-1}(1-\tilde{p})$ for $k\in\{1,2,\ldots\}$.  Thus, {the Bernoulli stick-breaking process has the following equivalent representation:}
\begin{equation}\label{eq:qb0}
	\begin{aligned}
		K & \sim \text{Geometric}(1-\tilde{p}),\\
		v_1,\dots,v_{K-1} & \sim \text{Beta}(1, \alpha), \quad v_K = 1, \\
		w_1 & = v_1, \quad w_k = v_k\prod_{l=1}^{k-1} (1-v_l), \quad k\geq 2.
	\end{aligned}
\end{equation}
The following result establishes the exchangeable partition probability function for {the Bernoulli stick-breaking process.}
\begin{lemma}\label{lemma:eppf0}
	The probability mass function of the random partition $\mathcal{A} = \{ A_1,\dots,A_t \}$ induced by $c$ in the Bernoulli stick-breaking process is
	\be
	\Pr_{\epsilon=0,n}(\mathcal{A}) = \frac{\alpha^t\Gamma(\alpha)}{\Gamma(n+\alpha)} \biggl(\prod_{j=1}^{t}\Gamma(n_j+1)\biggr)\sum_{{\sigma} \in S_t}
	\prod_{j=1}^{t}
	\frac{\tilde{p} + \mathds{1}(j=t) (1-\tilde{p}) / (\alpha \mathrm{B} (\alpha, n_{\sigma_t}+1)) } { g_j(\sigma) + \alpha (1 - \tilde{p}) }
	\ee
	where $n_j=|A_{j}|$, $g_j(\sigma) = \sum_{l=j}^t n_{\sigma_l}$, and {$\mathrm{B}(q_1,q_2)=\int_0^1 z^{q_1-1}(1-z)^{q_2-1}\,dz$, the Beta function.}
\end{lemma}

We now establish that {the Bernoulli stick-breaking process mixture model (quasi-Bernoulli mixture model with $\epsilon=0$)} exhibits posterior consistency for $K$ (the number of non-zero $w_k$'s) and $T$ (the number of clusters). {To clarify, even in the asymptotic case, $T$ could be less than the number of components $K$, if some component does not have any data point assigned to it.}
Let $\Omega$ denote the set of parameter tuples $\phi := (k, {w}_1,\dots,w_k,{\theta}_1,\dots,\theta_k)$ such that $k\in\{1,2,\ldots\}$, $w_1,\ldots,w_k> 0$, $\sum_{l=1}^k w_l = 1$, and $\theta_1,\dots,\theta_k \in \Theta$, the parameter space. Let $\Omega'$ denote the subset of $\Omega$ such that $\theta_i\neq \theta_j$ for all $i\neq j$. Further, let $P_{\phi}$ denote the mixture distribution $P_{\phi}:=\sum_{l=1}^{k} w_l \mathcal{F}(\theta_l)$.  When $\phi\in\Omega'$ is identifiable from $P_{\phi}$ up to permutation of the mixture components, we can define a transformation $\eta:\Omega\to \Omega'$ such that the parameter $\phi' = \eta(\phi)$ is fully identifiable from $P_{\phi'}$. See \citet[Section~3.2]{nobile1994bayesian} for the details. Any prior on $\Omega$ naturally defines an induced prior on $\Omega'$ through $\eta$.

\begin{theorem}\label{thm:consistent0}
	Assume $\phi\in \Omega'$ is identifiable up to permutation of the mixture components.
	Let $\Pi_0$ be the prior on $\Omega$ under the model defined by \cref{eq:qb0,eq:clusters}, and assume $\Pi_0(\{\phi:\exists\ i\neq j \text{ such that } \theta_i=\theta_j\})=0$. Let $\Pi_0'$ be the corresponding prior on $\Omega'$ induced by $\eta$.
	Then there is a subset $\Omega'_0 \subset \Omega'$ with $\Pi'_0(\Omega'_0) = 1$ such that for any $\phi_0 = (k_0,w^0_1,\ldots,w^0_{k_0},\theta^0_1,\ldots,\theta^0_{k_0}) \in \Omega'_0$,
	if $y_1,y_2,\ldots \mid \phi_0 \stackrel{iid}\sim P_{\phi_0}$
	%the support of the prior $\mathcal{G}(\theta)$ contains a neighborhood of each $\theta^0_k$
	and the component density $f_\theta$ is continuous (with respect to $\theta$) at each $\theta^0_k$, 
	then as $n\to\infty$, we have
	\be
	\Pr_{\epsilon=0}(K = k_0 \mid y_{1:n}) &\to 1 \quad \mathrm{a.s.}[P_{\phi_0}],\\
	\Pr_{\epsilon=0}(T = k_0 \mid y_{1:n}) &\to 1 \quad \mathrm{a.s.}[P_{\phi_0}],
	\ee
	{where $\mathrm{a.s.}[P_{\phi_0}]$ denotes almost surely convergence  under the probability distribution $P_{\phi_0}$.}
\end{theorem}
The first part of the result is a corollary of \citet[Proposition~3.5]{nobile1994bayesian}, the proof of which is an application of the Doob's theorem.  The intuition for the second part is that since $w^0_1,\ldots,w^0_{k_0}$ are positive and do not change with $n$, we can expect that at least some data will be assigned to each component $k = 1,\ldots,k_0$, and thus that $K$ and $T$ would match in the posterior.

Now, consider the case of $\epsilon>0$.  Intuitively, when $\epsilon$ is small, we would expect the posterior to behave similarly to the case of $\epsilon = 0$.  Formally, we show that this is indeed the case when $\epsilon=\epsilon(n) \to 0$ at an appropriate rate as $n\to\infty$. To show this, we employ the following bound on the distance between the partition distributions for $\epsilon>0$ and $\epsilon=0$ under the prior.

\begin{theorem}[Prior equivalence as $\epsilon(n) \to 0$]\label{thm:prior}
	Assume $\epsilon \leq 1/n$. Under the prior, the total variation distance between the partition distributions for $\epsilon > 0$ and $\epsilon = 0$ satisfies the bound
	\be
	\sup_{A \in \mathscr A} | \Pr_{\epsilon,n}(A) -  \Pr_{0,n}(A)| \leq \sqrt {\frac{\alpha n\epsilon}{2(\alpha+1-\alpha\epsilon n)}} 
	\ee
	where $\mathscr{A}$ denotes the set of all subsets of $\cup_{t=1}^n \mathcal{H}_t(n)$, and  $\Pr_{\epsilon,n}(A) = \sum_{\mathcal{A}\in A} \Pr_{\epsilon,n}(\mathcal{A})$.
	In particular, if $\epsilon(n) = o(1/n)$, then 
	\be
	\sup_{A \in \mathscr A} | \Pr_{\epsilon(n),n}(A) -  \Pr_{0,n}(A)| \xrightarrow[n\to\infty]{} 0.
	\ee 
\end{theorem}

The interpretation of this result is that if we control the tail to be slightly smaller than $1/n$, then we have a stick-breaking model supported in the infinite-dimensional space that asymptotically approximates the finite-dimensional model with posterior consistency guarantees. 

Now, moving to the posterior, we have the consistency of the  quasi-Bernoulli model for the number of clusters.

\begin{theorem}[Posterior consistency]\label{thm:consistent}
	%Let $\Omega$ be the same as in \cref{thm:consistent0}, and $\Pi_0$ be the prior on $\Omega$ from the quasi-Bernoulli model using \cref{eq:qbcluster} with $\epsilon>0$. 
	Under the same assumptions and notations of \cref{thm:consistent0}, 
	%there is a subset $\Omega'_0 \subset \Omega'$ with $\Pi'_0(\Omega'_0) = 1$ such that , 
	if $\epsilon(n) = o(1/n^{2+r})$, where $r$ is the integer such that $\max(\alpha-1,0)\le r< \alpha$,
	%is the unique integer such that $0<\alpha - r \leq 1$, 
	then
	\begin{equation*}
		\Pr_{\epsilon(n)}(T = k_0 \mid y_{1:n}) \xrightarrow[n\to\infty]{} 1 \quad \mathrm{a.s.}[P_{\phi_0}].
	\end{equation*}
\end{theorem}
Note that here we assume $\epsilon(n) = o(1/n^{2+r})$ rather than $o(1/n)$ as in \cref{thm:prior}, however, we expect that the involved inequalities could be tightened further.

It should also be noted that letting $\epsilon(n)$ depend on $n$ makes the resulting sequence of models  no longer projective.  That is, the model for $n$ data points does not coincide with the distribution obtained by taking the model for $n+1$ data points and integrating over data point $n+1$. However, in order to achieve certain optimal asymptotic behaviors such as consistency, it is common to calibrate the prior based on the sample size (for example, see \cite{castillo2015bayesian} on the choice of prior for variable selection). Alternatively, one could always use $\epsilon=0$, which achieves both consistency and projectivity, but this is less flexible and less efficient in terms of computation.

\subsection{Comparison with the Asymptotic Behavior of the Dirichlet Process}\label{subsec:comparedp}

The results of \cite{miller2013simple,miller2014inconsistency} on the inconsistency of the Dirichlet process mixture model show that for a fixed value of the concentration parameter $\alpha$, the model asymptotically over-estimates the number of clusters on data from a finite mixture.

Our theoretic finding in the quasi-Bernoulli raises a tempting question: can we achieve consistency with a Dirichlet process mixture if we let $\alpha=\alpha(n)$ go to $0$ at an appropriate rate as $n$ increases? To our knowledge, this remains an open question. However, here we provide a partial answer (in the negative), by showing that if $\alpha(n) \to 0$ too fast, then the Dirichlet process remains inconsistent for the number of clusters.

\begin{lemma}\label{lemma:dp0_inconsistency}
	Suppose the data are $y_1,\dots,y_n \stackrel{iid}\sim  0.5 \No(0,1)+0.5\No(\kappa,1)$  for any fixed $\kappa\in\mathbb{R}$, {where $\No(\mu,\sigma^2)$ is the Gaussian distribution with density $f_{\mu,\sigma^2}(x) \propto \exp{\{-(x-\mu)^2/(2\sigma^2)\}}$.}  Consider a Dirichlet process mixture model {with the mixture components $\mathcal{F(\theta)}=\No(\theta, 1)$}, the base measure $\mathcal{G} = \No(0,1)$ {(the prior on the parameter $\theta$)}, and concentration parameter $\alpha = \alpha(n)$ such that $\alpha(n) = o(\exp(-a_0\,n))$ for some constant $a_0 > 1/2 + \kappa^2/4$.  Then $\Pr(T=2 \mid y_{1:n}) \xrightarrow{\mathrm{a.s.}} 0$ as $n\to\infty$.
\end{lemma}
{
	To clarify, the purpose of the above result is not to provide a practical guidance on choosing the rate of $\alpha(n)\to 0$.  Indeed, the problem of the Dirichlet process mixture is usually overestimation rather underestimation of the number of clusters in the limit \citep{yang2019posterior}. 
	Rather, this result shows that if consistency could be achieved for the Dirichlet process mixture model under a sample-size-dependent $\alpha(n)$, the rate of this hyper-parameter needs to satisfy both an upper and a lower bounds, which may turn out to be practically challenging for the users --- there is a somewhat delicate sensitivity issue. In comparison, a strength of the quasi-Bernoulli mixture is that for obtaining consistency $\epsilon$ only needs to satisfy one upper bound $o(1/n^{2+\alpha})$, which means the user can simply use a small $\epsilon$ such as $1/n^{2+\alpha}$ (or smaller), without worrying about impacting the consistency. We provide empirical comparison via simulations with different cases of $\alpha(n)$ and $\epsilon(n)$ in the \cref{subsec:dp_sim}.

	The above result does not exclude the possibility that the Dirichlet process with $\alpha\to 0$ at a slower rate could achieve consistency. For example, recently \cite{ohn2020optimal} show that setting $\alpha \approx {n^{-a_1}}$ for some positive number $a_1$ will guarantee $\t{pr}(T > C k_0 \mid y_{1:n})\to 0$ for some constant  $C>1$, hence preventing severe over-estimation in the number of clusters --- although whether one could exactly recover $k_0$ is still unknown. Alternatively, another possibility is to put a hyper-prior on $\alpha$. \cite{ascolani2022clustering} show that this method can achieve consistency when the component distribution $\mathcal{F}$ has bounded support. To our best knowledge, the consistency with general $\mathcal{F}$ still remains an open question.}

\section{Posterior Sampling Algorithm}\label{sec:sampling}

Since the quasi-Bernoulli mixture model involves a small modification to classic stick-breaking construction, we can use {an efficient slice sampling algorithm [\cite{kalli2011slice}, as the improved version of \cite{walker2007sampling}] for posterior inference. We use  {a sequence of decreasing positive constants $\xi_1,\xi_2,\ldots$ that converges to zero}. {In this article, we choose $\xi_i =  0.5^{i}$ for $i\geq 1$ as suggested in \cite{kalli2011slice}.} 
	Given $c_i$, consider a latent uniform $u_i \sim \text{Uniform}(0,\xi_{c_i})$, then we have a joint likelihood proportional to
	\(
	\prod_{i=1}^n \mathds{1}(u_i<\xi_{c_i})w_{c_i}/\xi_{c_i}f_{\theta_{c_i}}(y_i)
	\).
}
We define the state of the Markov chain to be $(c,\theta,w,u)$ and the target distribution is the posterior
$p(c,\theta,w,u \mid y)$, where $y=y_{1:n}$, $c=c_{1:n}$, $\theta=\theta_{1:\infty}$, and $w=w_{1:\infty}$.

The slice sampler iterates the following steps:
\begin{enumerate}
	\item \textit{Sample $c$ from its full conditional.} For $i=1,\dots,n$: sample $c_i \sim \t{Categorical}(\tilde{w})$ where
	{
		\[
		\tilde{w}_k = \frac{w_k/\xi_k f_{\theta_k}(y_i)}{ \sum_{\{l:\xi_l>u_i\}} w_l/\xi_l f_{\theta_l}(y_i)  }, \quad k\in \{l:\xi_l>u_i\}.
		\]
		Since the sequence $\xi_1,\xi_2,\dots$ converges to zero, the index set $\{l:\xi_l>u_i\}$ is finite.}
	Compute $n_k := \sum_i \mathds{1}(c_i=k)$, and $m_k := \sum_i \mathds{1}(c_i>k)$.
	\item {\textit{Sample $u$ from its full conditional.} For $i=1,\dots,n$: sample $u_i$ from the uniform distribution over the interval $(0,\xi_{c_i})$.}
	\item \textit{Sample $w$ from its full conditional.} For {$k\in \cup_{i=1}^n\{l:\xi_l>u_i\}$}: 
	\begin{itemize}
		\item Sample $b_k\sim q\delta_1(\cdot) + (1-q)\delta_\epsilon(\cdot)$ where
		\[
		q = \frac{\tilde{p}}{\tilde{p} + (1-\tilde{p}) \epsilon ^ {-\alpha} I_\epsilon(m_k+\alpha,n_k+1)}.
		\]
		\item Sample $\beta_k$ by drawing $X\sim \text{Beta}_{(0,b_k)} (m_k + \alpha, n_k+1)$ and setting $\beta_k = X/b_k$,
		where $\text{Beta}_{(0,\epsilon)}$ denotes a Beta distribution truncated to the interval $(0,\epsilon)$.
		\item Compute $w_k$ from $b_{1:k}$ and $\beta_{1:k}$ using \cref{eq:qbsb}.
	\end{itemize}   
	\item \textit{Sample $\theta$ from its full conditional.} For $k\in \cup_{i=1}^n\{l:\xi_l>u_i\}$: sample $\theta_k$ from the distribution proportional to $g(\theta_k)\prod_{i:c_i=k} f_{\theta_k}(y_i)$, {where $g$ is the density of the base measure $\mathcal{G}$}.
\end{enumerate}
Here, $f_\theta$ denotes the density of the component distribution $\mathcal{F}(\theta)$.
Note that in the $w$ update, we first sample $b_k$ marginalized over $\beta_k$, then sample $\beta_k \mid b_k$,
and compute the resulting value of $w_k$. 

\section{Simulations}\label{sec:sim}

In this section, we assess the empirical performance of the quasi-Bernoulli (QB) mixture model in simulation studies. We compare with three popular alternatives: Dirichlet process (DP) mixture, Pitman--Yor process (PY) mixture and finite mixture with a prior on the number of components (MFM).

We demonstrate the consistency of the quasi-Bernoulli mixture model for the number of clusters $T$ when the family of component distributions is correctly specified.
We set $\tilde{p} = 0.9$ for the quasi-Bernoulli probability in \cref{eq:qbsb}, yielding a prior mean of {no more than $1/(1-\tilde{p}) = 10$ components with mixture weights larger than $\epsilon$. The reason is that the random variable $K$ for the first time $b_k=\epsilon$ follows Geometric($1-\tilde{p}$), and the weights after $K$ are less than $\epsilon$.}
Alternatively, one could place a Beta hyper-prior on $\tilde{p}$ to make the prior even more weakly informative. We set $\alpha=1$ and $\epsilon = 1/n^{2.1}$, which satisfies the theoretical condition that $\epsilon = o(1/n^2)$ in \cref{thm:consistent}. 
To have a fair comparison, we set the Dirichlet process concentration parameter $\alpha$ so that the expected number of clusters under the Dirichlet process prior is as close as possible to the one under the quasi-Bernoulli process prior for each $n$, as shown in \cref{tab:hyper-parameter} in the appendix.
Similarly, for the Pitman--Yor process prior ${\text{PY}}(\alpha,d)$ where each $v_k$ follows $\text{Beta}(1-d,\alpha+kd)$, we choose $\alpha$ and $d$ to match both the expectation and the variance of number of clusters with quasi-Bernoulli process prior.
For the MFM model, we follow \cite{miller2018mixture} and set $\Pi_K(k)=p^{k-1}(1-p)$ where $p=0.9$ for the prior on number of components, and $(w_1,\dots,w_K)\sim \Dir_K(\alpha,\dots,\alpha)$ where $\alpha=1$ for the prior on mixture weights. 

For each experiment, we run the Markov chain for 50,000 iterations, discard the first 20,000 as burn-ins, and use thinning by keeping only every 50th iteration.% \cheng{We find this is enough by checking the trace plots and the $\hat{R}$ statistic \citep{gelman1992inference}.}

\subsection{Simulations with Gaussian Mixtures}\label{subsec:uniGau}

To compare performance in terms of consistency and MCMC mixing, we first consider simulations using Gaussian distributions $\No(\mu,\Sigma)$ for the mixture components.

We first generate data with sample sizes $n\in\{50,100,250,1000,2500\}$ from a three-component univariate Gaussian mixture distribution: $0.3\No(-4, 1^2) + 0.3\No(0, 1^2) + 0.4\No(5, 1^2)$. Following \cite{richardson1997bayesian}, we use a data-dependent prior (that is, base measure $\mathcal{G}$) on the component parameters $(\mu,\Sigma)$: $\mu \sim \No(m_\mu,s^2_\mu)$ and $\Sigma \sim \Ga^{-1}(2,\gamma)$
where $\Ga^{-1}(a, b)$ has density $f(x)\propto x^{-a-1}\exp(-b/x)$, with a hyper-prior $\Ga(g,h)$ on $\gamma$, where $m_\mu = (\max\{y_{1:n}\} + \min\{y_{1:n}\})/ 2,\ s_\mu = \max\{y_{1:n}\} - \min\{y_{1:n}\}$, $g = 0.2$, and $h = 10 / s_\mu^2$.  

%We report the resulting posterior probabilities of $T=t$.

\begin{figure}
	\centering
	\includegraphics[width=1\linewidth]{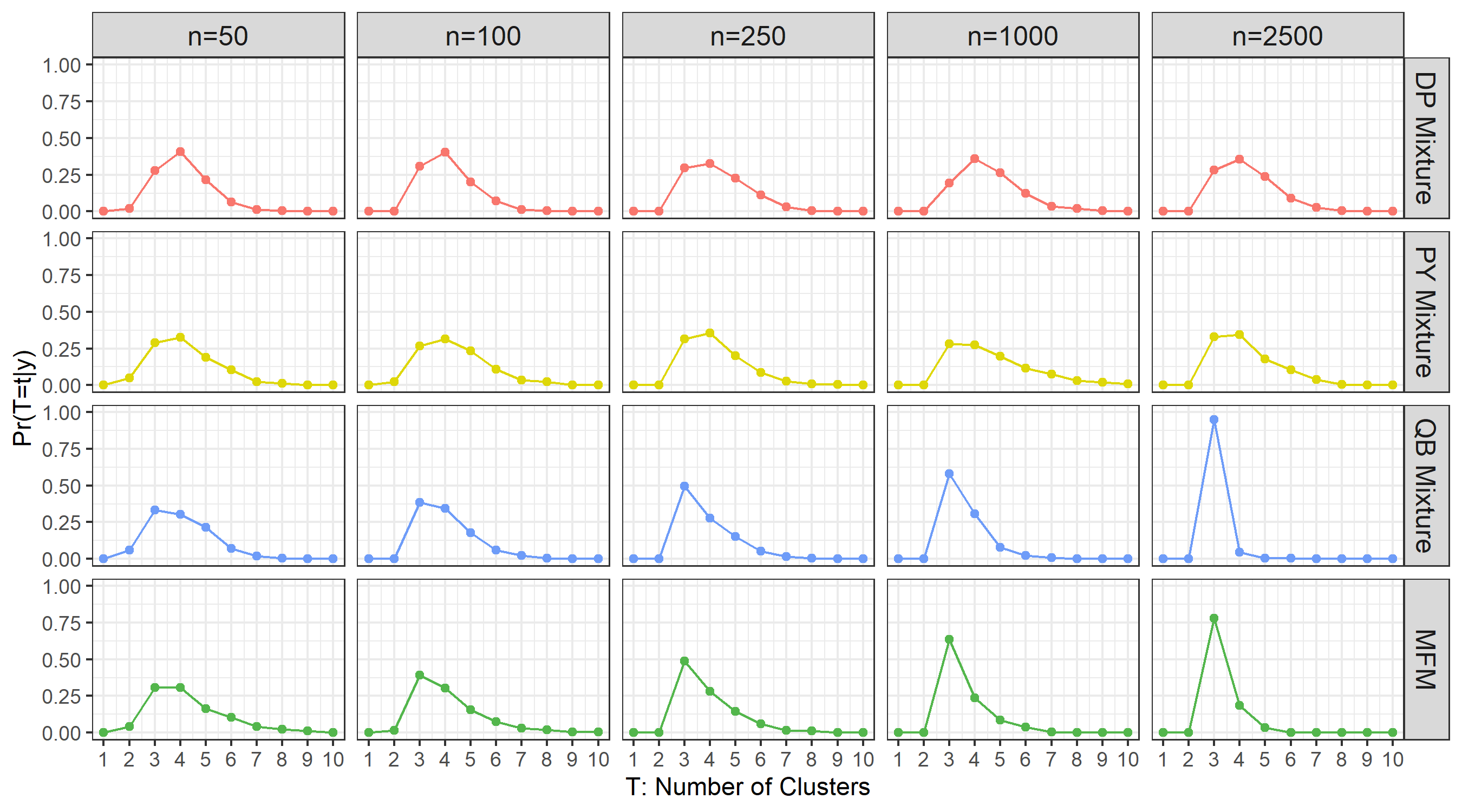}
	\caption{Posterior distribution of the number of clusters ($T$) for data from a three-component univariate Gaussian mixture.
		The quasi-Bernoulli mixture model correctly concentrates on three clusters, and its posterior distribution of $T$ converges to a point mass at $k_0=3$. {Coherent with our theory, at large $n$, the posterior distributions become almost identical to the ones using the MFM model.} However, the posterior distributions of calibrated DP mixture model and PY mixture model do not converge to a point mass at $k_0=3$.} \label{fig:1D}
\end{figure}

\Cref{fig:1D} plots the posterior distribution of the number of clusters $T$ at each $n$. Under the quasi-Bernoulli mixture model (shown in blue), the posterior of $T$ converges to a point mass at the true number of components ($k_0 = 3$) as $n$ grows, in accordance with our theory. Further, for both small and large $n$, the posterior mode of $T$ coincides with the true number of components. {Clearly, our model yields almost the same results (blue) as the MFM model (green), especially at large $n$. This is coherent with our theory (\cref{eq:qb0}).} On the other hand, the Dirichlet process mixture model and the Pitman--Yor process mixture model fail to concentrate at the true number of components.

\begin{figure}
	\begin{subfigure}[t]{0.49\textwidth}
		\centering
		\includegraphics[width=1\linewidth]{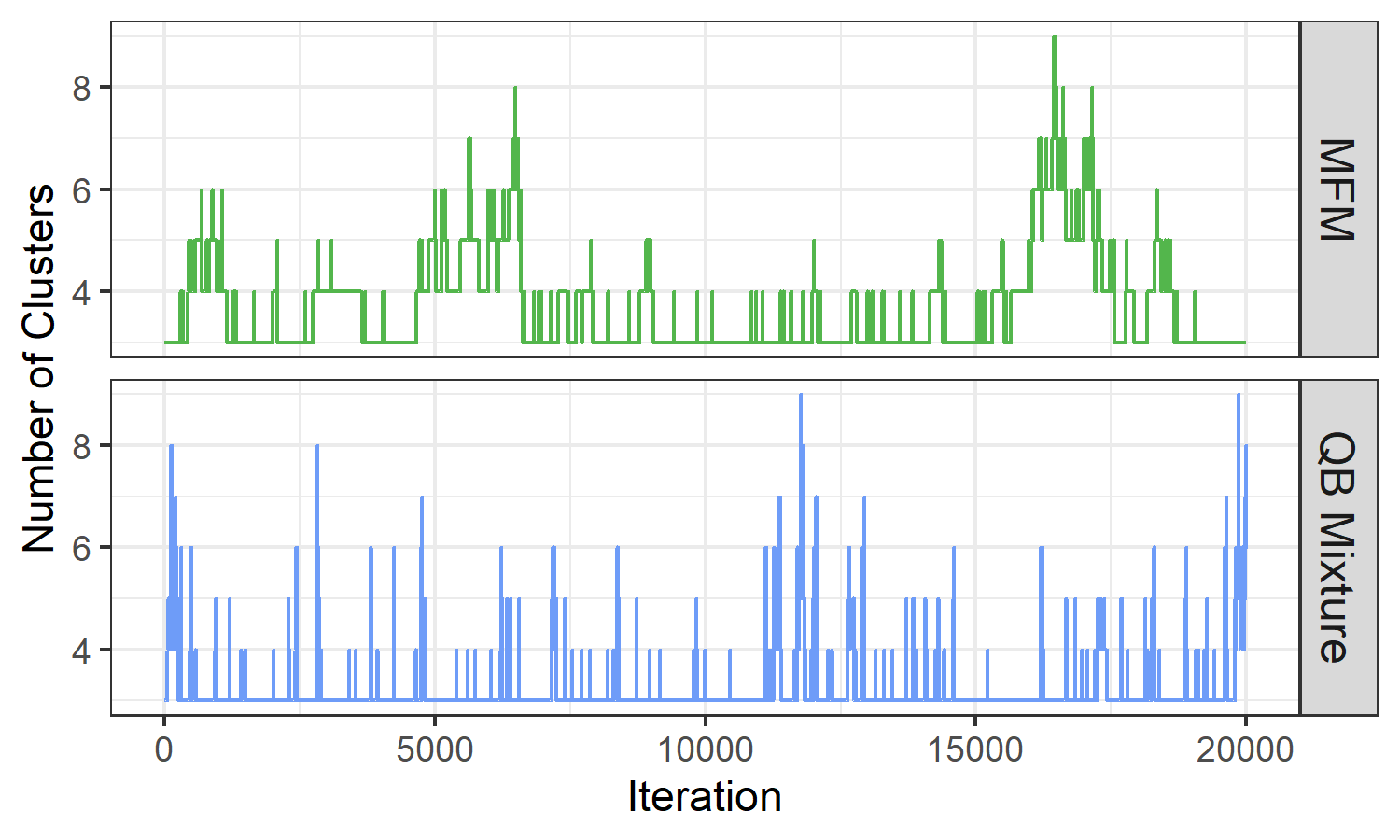}
		\caption{Trace on number of clusters}
	\end{subfigure}
	\begin{subfigure}[t]{0.49\textwidth}
		\centering
		\includegraphics[width=1\linewidth]{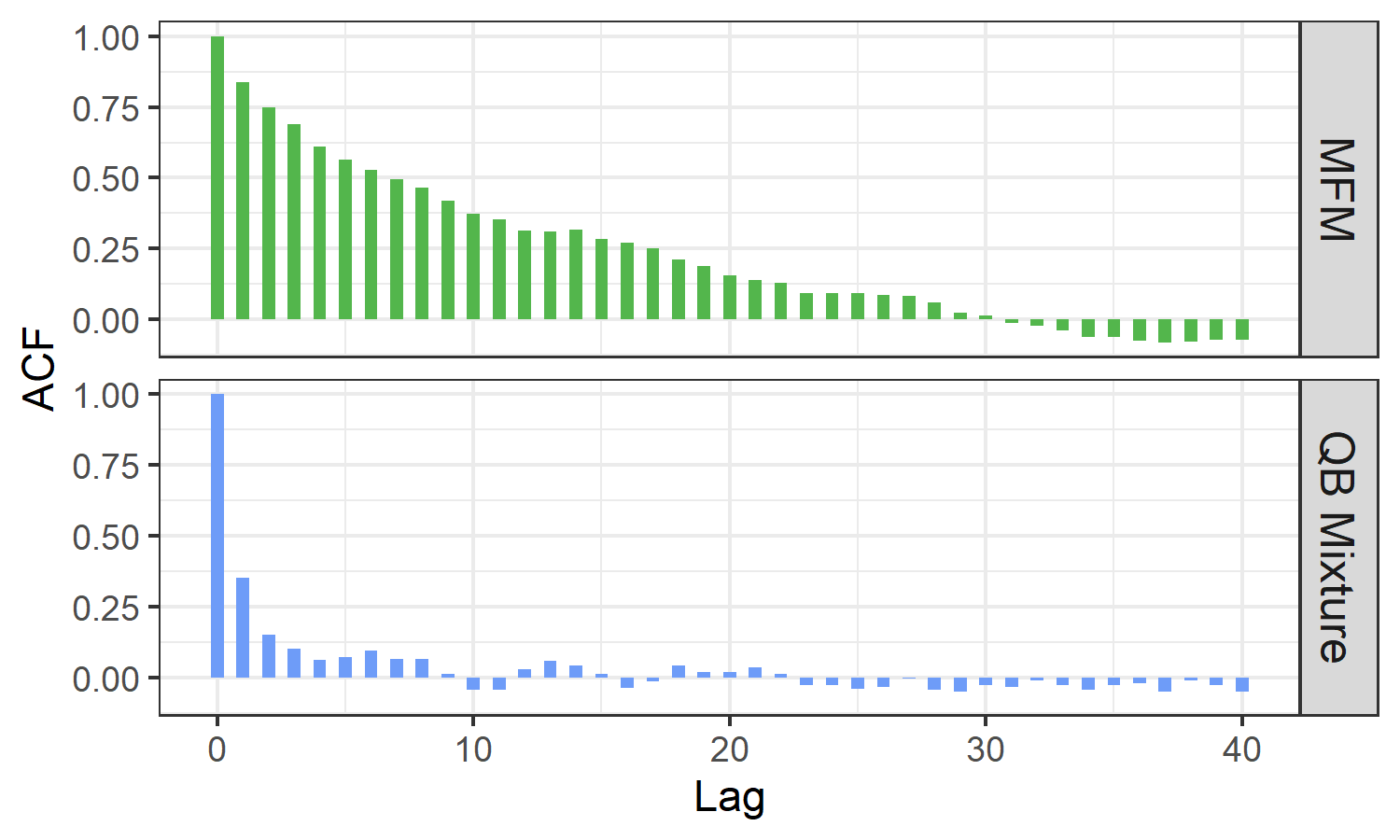}
		\caption{Auto-correlation (based on thinning at $50$)} 
	\end{subfigure}
	\centering
	\caption{The trace of the Markov chain on \(T\) and auto-correlation functions for univariate Gaussian mixture data with sample size \(1000\). Quasi-Bernoulli mixture model shows much better mixing in the Markov chain, compared to the MFM model. We discard the first 5,000 iterations as burn-ins and record the following 20,000 samples.}
	\label{fig:computing_perf}
\end{figure}

{Despite similar performances in achieving consistency, a major strength of our model is its computational efficiency gained via the slice sampling \citep{kalli2011slice}. In comparison, the existing MFM model requires a combinatorial search via the split-merge sampler \citep{jain2007splitting}, which suffers from slow mixing with high auto-correlation.
	As shown in Figure~\ref{fig:computing_perf}, with thinning, quasi-Bernoulli mixture model using slice sampler shows a drop in the auto-correlation  (effective sample size $16.0\%$, on average of five experiments with sample size $1000$), while MFM model using the split-merge sampler shows a much slower drop (effective sample size $7.8\%$).
}

Next, we consider a multivariate simulation scenario in which we generate data sets of size $n\in\{250, 1000, 2500\}$ from a three-component bivariate Gaussian mixture: $0.3\No((-4\ 1)^{\rm T}, I_2) + 0.3\No((0\ 2)^{\rm T}, I_2) + 0.4\No((5\ 3)^{\rm T}, I_2)$. We use the data-dependent prior $\mu \sim \No(m, C)$, $\Sigma \sim \Wishart^{-1}_2(C^{-1}/2,2)$ on the component parameters, where $m$ is the sample mean and $C$ is the sample covariance.
The results are similar to the univariate simulation scenario; see \cref{subsec:sim_multiGau}.

\subsection{Simulations with Non-Gaussian Mixtures}

The quasi-Bernoulli mixture model can easily be extended to mixture models with non-Gaussian components. To illustrate that the consistency result still holds, we consider data generated from a mixture of Laplace distributions. 

\begin{figure}
	\centering
	\includegraphics[width=1\linewidth]{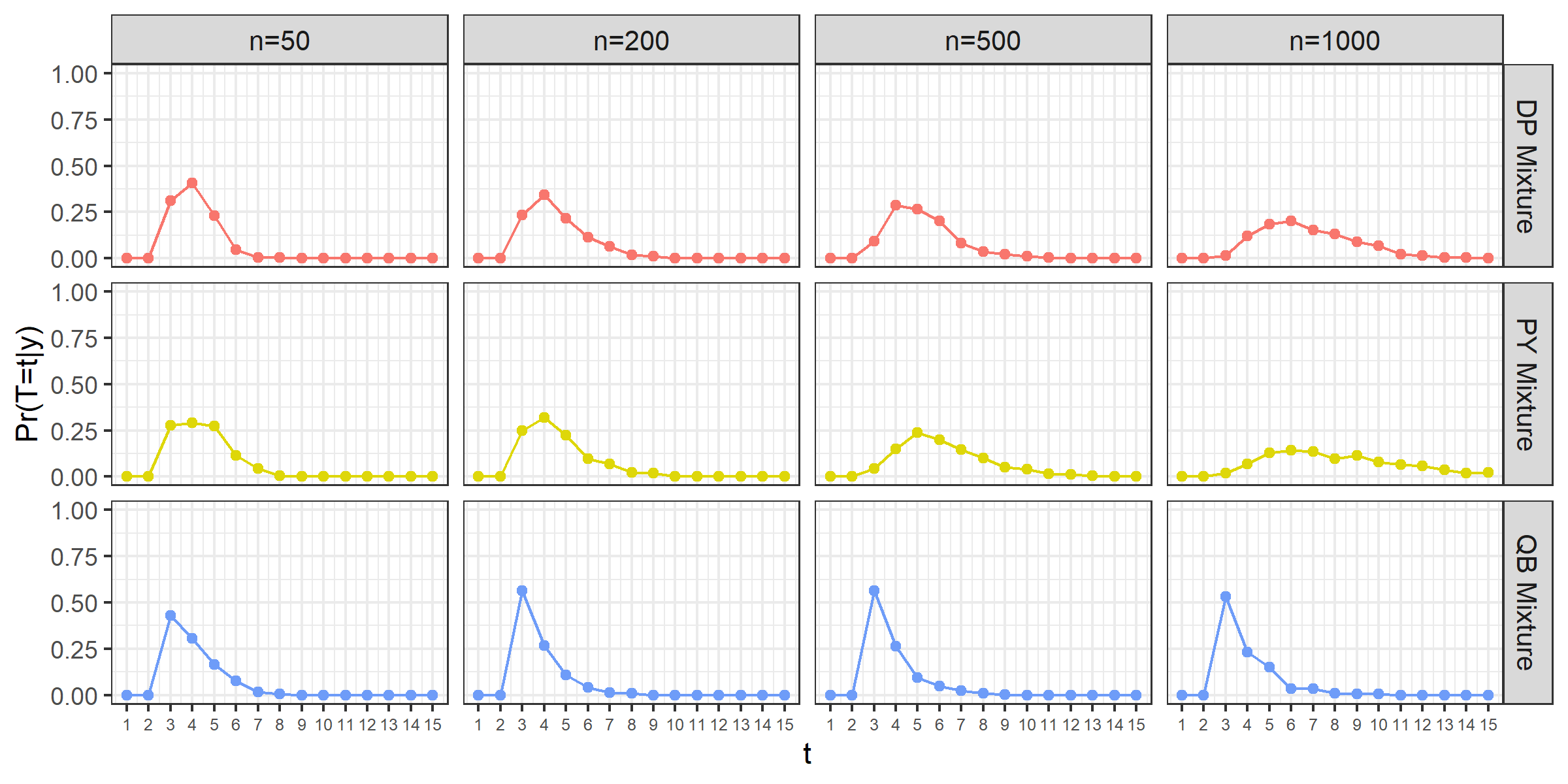}
	\caption{Quasi-Bernoulli mixture model correctly recovers three clusters as the ground truth when each component is from a Laplace distribution.  The Dirichlet process and Pitman--Yor process over-estimate the number of clusters, due to having small spurious clusters. \label{fig:laplace}}
\end{figure}
We generate data from a three-component Laplace mixture: $0.35\t{Lap}(-10,1) + 0.3\t{Lap}(0, 1.5) + 0.35 \t{Lap}(10, 0.5)$, where $\t{Lap}(\mu, \lambda) $ denotes a Laplace distribution with mean $\mu$ and scale $\lambda$. We use a data-dependent prior (base measure $\mathcal{G}$) on $(\mu,\lambda)$: $\mu \sim \No(m_\mu,\sigma^2_\mu)$ and $\lambda \sim \Ga^{-1}(2,1)$, where $m_\mu = (\max\{{y_{1:n}}\} + \min\{{y_{1:n}}\} ) / 2$ and $\sigma_\mu = \max\{{y_{1:n}}\} - \min\{{y_{1:n}}\}$. \Cref{fig:laplace} shows that the quasi-Bernoulli process successfully recovers the true number of components, while the Dirichlet process (red) and the Pitman--Yor process (yellow) fail to do so. Under a mixture distribution like the Laplace distribution having heavier tail than the Gaussian distribution, the Dirichlet process and the Pitman--Yor process tend to over-estimate the number of clusters to a greater extent.

\section{Data Application: Clustering Brain Networks}
To demonstrate the ease of using our model in an advanced data analysis, we apply it to cluster multiple brain networks, collected from  $n=812$ subjects in the human connectome project \citep{marcus2011informatics}.

For each subject, resting-state functional magnetic resonance imaging (fMRI)  signals were collected from $R=50$ regions of the brain, indexed by $r=1,\ldots,R$. The data were processed and transformed into a connectivity graph on $R$ vertices, represented as a symmetric binary adjacency matrix $Y^{(i)}\in \{0,1\}^{R\times R}$ such that $Y^{(i)}=Y^{(i)\rm T}$.

We model these adjacency matrices using a probit-Bernoulli mixture model in which each component distribution has low-rank latent structure. {The goal of this model is to cluster the networks and to find meaningful low-dimensional smoothing of networks, by borrowing information within each sub-group/cluster. }

Given a matrix of probabilities $\theta\in[0,1]^{R\times R}$, we write $Y\sim \t{Bernoulli}(\theta)$ to denote that $Y_{r s}\sim \t{Bernoulli}(\theta_{r s})$ independently for $r,s\in\{1,\ldots,R\}$.  In this notation, we use the following infinite mixture model with a quasi-Bernoulli stick-breaking prior on the mixture weights $w = (w_1,w_2,\ldots)$,
\bel\label{eq:network_model}
Y^{(i)} \mid c_i,\mu,M & \sim \t{Bernoulli}(\Phi(\mu + M_{c_i})) \text{ independently for } i=1,\ldots,n, \\
c_1,\ldots,c_n\mid w & \stackrel{iid}\sim \t{Categorical}(w),  \\
M_k & = Q_k \Lambda_k Q_k^{\rm T}, \quad k\geq 1,\\
w & \sim \text{Quasi-Bernoulli}(\tilde p,\epsilon,\alpha),
\eel
where $\Phi(\cdot)$ is the cumulative distribution function of the standard Gaussian distribution, applied element-wise, {and the weights $w = (w_1,w_2,\ldots)$ are drawn from \cref{eq:qbsb} which is denoted by Quasi-Bernoulli($\tilde p,\epsilon,\alpha$)}. {The model enforces symmetry for the binary matrices $Y^{(i)}$ by only modeling the lower triangle part of them.}
Here, $\mu$ is a scalar that is shared by all components, and each $M_k = Q_k \Lambda_k Q_k^{\rm T}$ is a component-specific matrix such that $\Lambda_k=\text{diag}(\lambda_{k,1},\ldots,\lambda_{k,d})$, where $\lambda_{k,l}>0$,
and $Q_k$ belongs to the Stiefel manifold $\mathcal{V}^{d,R}:=\{Q \in \mathbb{R}^{R\times d}: Q^{\rm T}Q=I_d\}$ \citep{hoff2009simulation}.
The role of $M_k$ is to provide a low-rank representation for mixture component $k$,
and the $\mu$ is a nuisance parameter that captures departures from this assumed low-rank structure.

%It is tempting to model each network as the collection of Bernoulli random variables, with the probability matrix directly parameterized by a group-specific latent matrix. However,  a dilemma in  clustering high-dimensional data is that the clustering tends to capture nuanced fluctuations, leading to excessively many clusters. This can lead to asymptotic failure of collecting $n$ clusters as dimension diverges, as discovered recently by \cite{chandra2020bayesian}.

%To avoid this issue, we propose to consider an {\em overdispersed} probit-Bernoulli mixture model, in the following form:
%\be
%    Y^{(i)} & \sim \t{Bernoulli}(\Phi(\mu + M_{c_i} + D^{(i)})) \t{ for } i=1,\ldots,n, \\
%    c_1,\ldots,c_n & \sim \t{Categorical}(w), \\
%    M_k & = Q_k \Lambda_k Q_k^{\rm T} \t{ for } k=1,2,\ldots,\\
%\ee
%where $\Phi(\cdot)$ is the cumulative distribution function of the standard Gaussian distribution, applied element-wise.
%Here, we model the group-specific structure via a low-rank matrix  $M_k = Q_k \Lambda_k Q_k^{\rm T}$, such that $\Lambda_k=\text{diag}(\lambda_{k,1},\ldots,\lambda_{k,d})$, where $\lambda_{k,l}>0$,
%and $Q_k$ belongs to the Stiefel manifold $\mathcal{V}^{d,R}:=\{Q \in \mathbb{R}^{R\times d}: Q^{\rm T}Q=I_d\}$ \citep{hoff2009simulation}. On the other hand, $D^{(i)}$ is a symmetric matrix that accommodates overdispersion (corresponding to perturbation from the group-specific structure, that are not able to be captured by the probit-Bernoulli alone), we assign each of its elements to a Gaussian prior $\t{N}(0, \sigma^2_D)$, with $\sigma^2_D \sim \t{IG}(1,1)$.

For the other priors, we assign $Q_k \sim \t{Uniform}(\mathcal{V}^{d,R})$ for $k=1,2,\ldots$ with $d = 2$, use a truncated Gaussian prior $\pi(\lambda_{k,l})\propto \No(\lambda_{k,l}\mid 0,50)\mathds{1}(\lambda_{k,l}>0)$ for $l=1,\ldots,d$, and assign a Gaussian prior $\No(0,10^2)$ on $\mu$, following \citet{hoff2009simulation}.
For the quasi-Bernoulli prior on $w$, we use $\tilde{p} = 0.9$, $\alpha=1$ and $\epsilon = 1 / n^{2.1}$. We run the MCMC sampler from \cref{sec:sampling} for $30,000$ iterations and discard the first $10,000$ as burn-ins.

\begin{figure}
	\centering
	\includegraphics[width=1\linewidth]{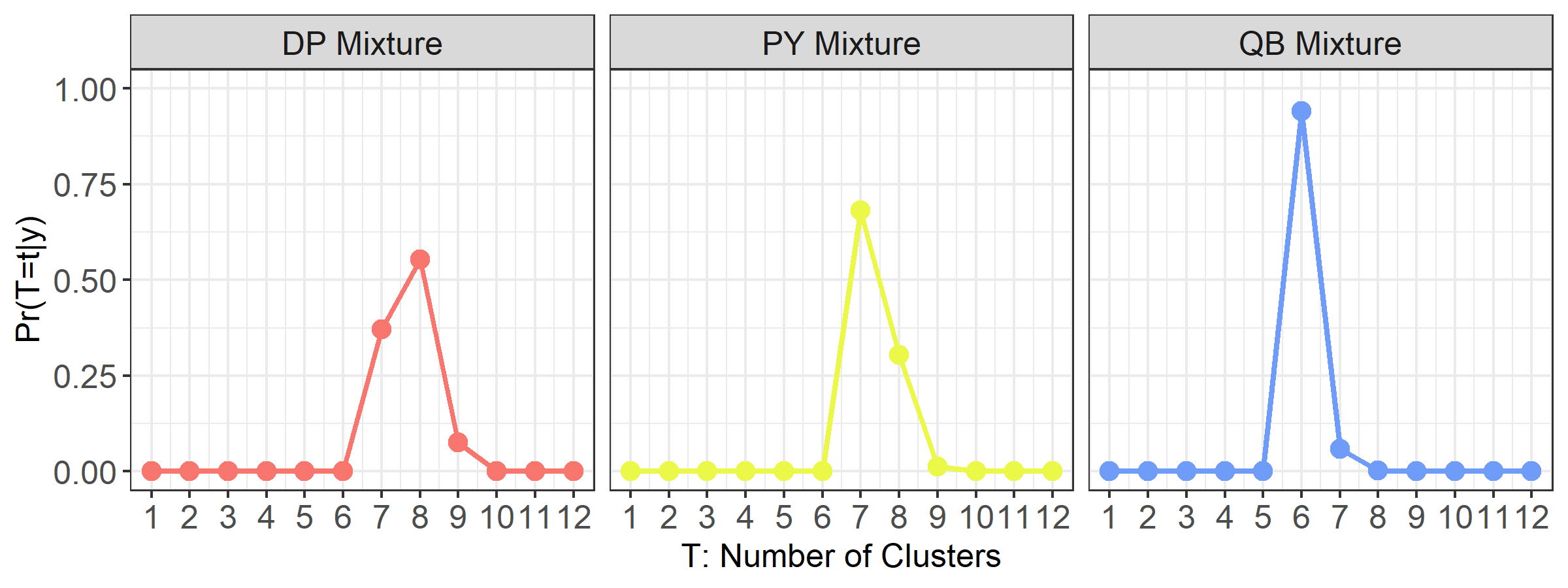}
	\caption{The quasi-Bernoulli model concentrates on $T=6$ clusters on the brain connectivity data. For comparison, the posterior mode of the corresponding Dirichlet process and Pitman--Yor process mixture models are at $T=8$ and $T=7$.\label{fig:fmri}}
\end{figure}

\begin{figure}
	\centering
	%\begin{subfigure}[t]{0.19\textwidth}
	%        \centering
	%        \includegraphics[width=1\linewidth]{fmri_legend.png}
	%\end{subfigure}\\
	\begin{subfigure}[t]{0.32\textwidth}
		\centering
		\includegraphics[width=1\linewidth]{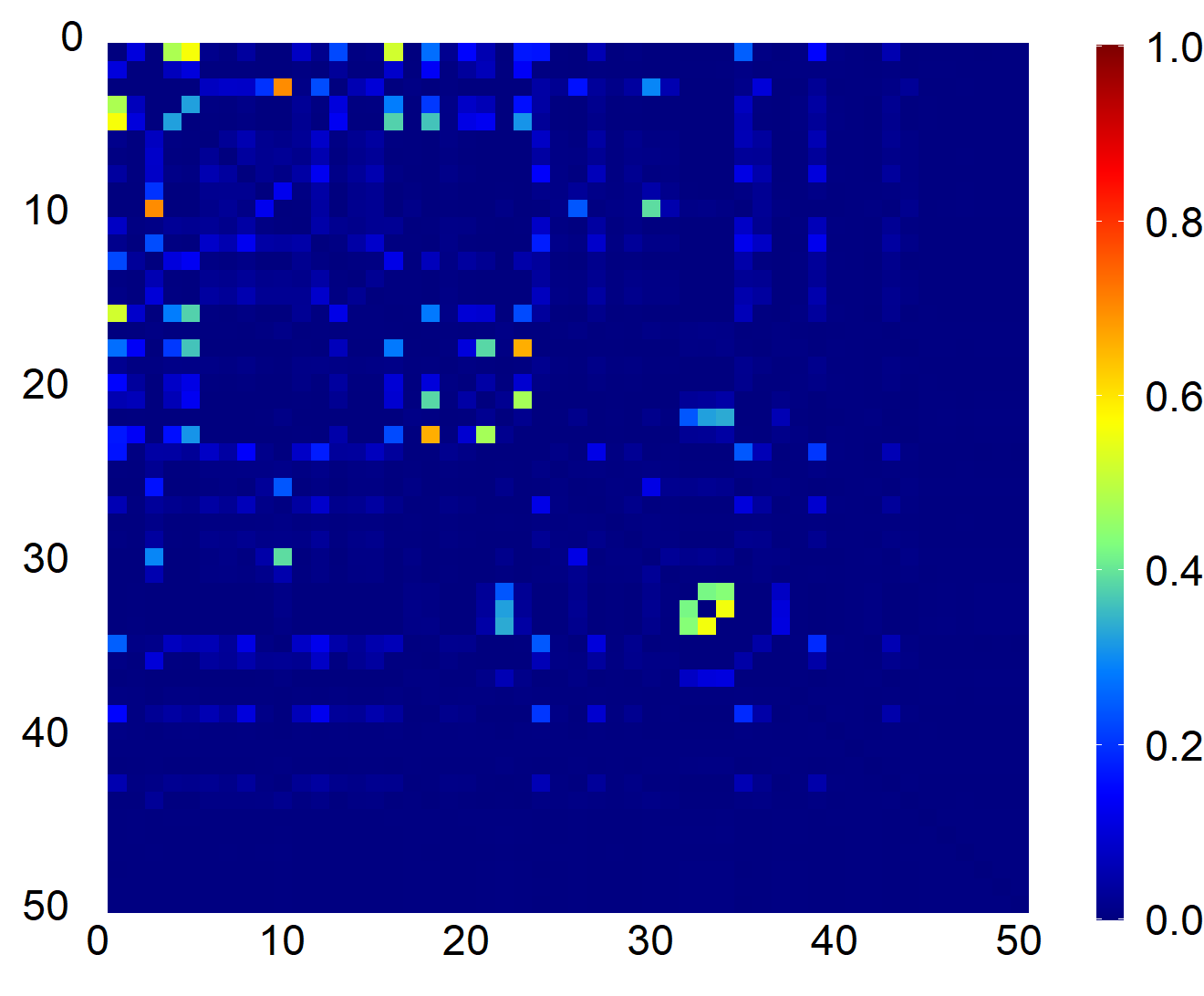}
		\caption{Group 1 ($46.0\%$ of the subjects)}
	\end{subfigure}
	\begin{subfigure}[t]{0.33\textwidth}
		\centering
		\includegraphics[width=0.97\linewidth]{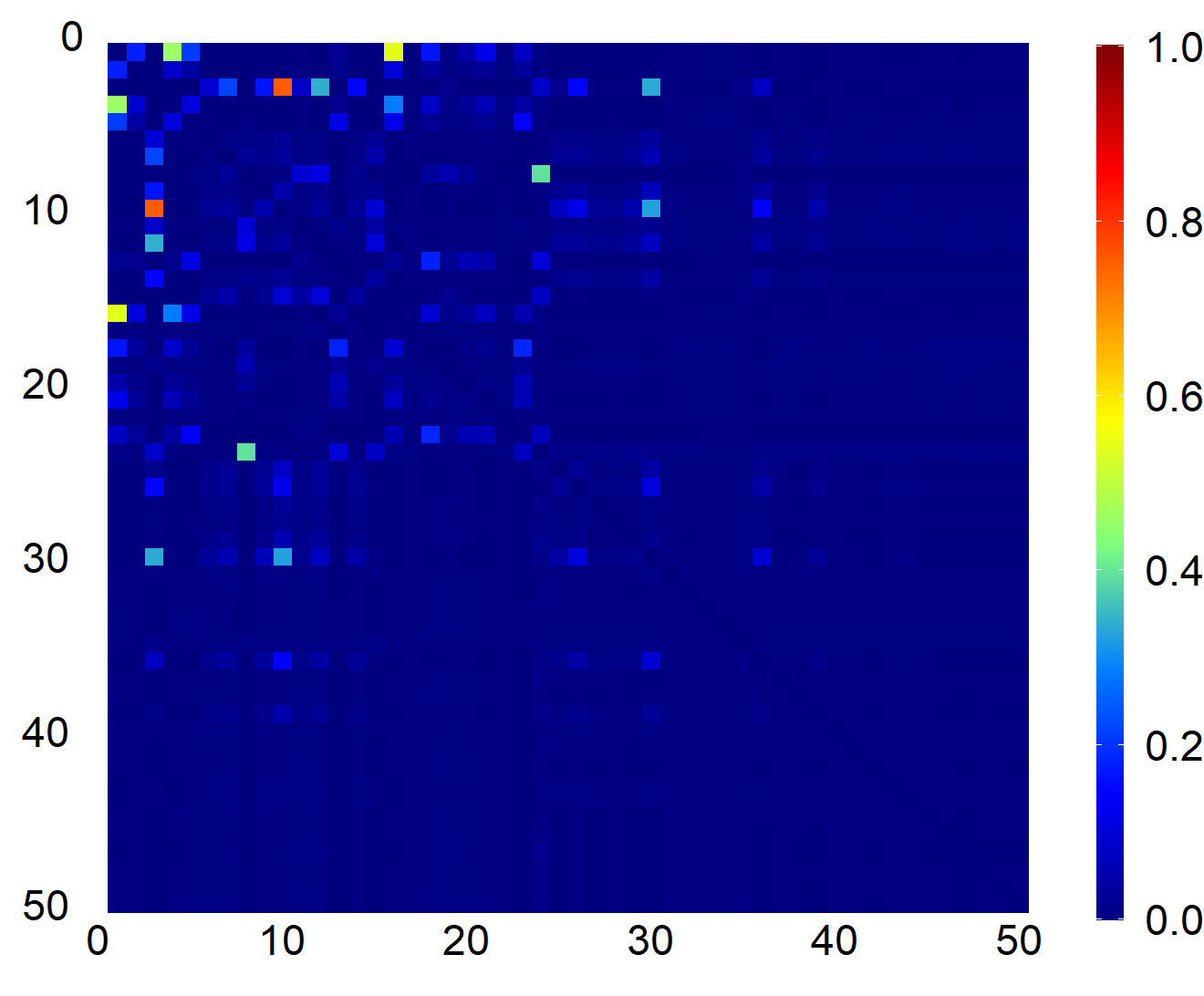}
		\caption{Group 2 ($18.9\%$ of the subjects)}
	\end{subfigure}
	\begin{subfigure}[t]{0.32\textwidth}
		\centering
		\includegraphics[width=1\linewidth]{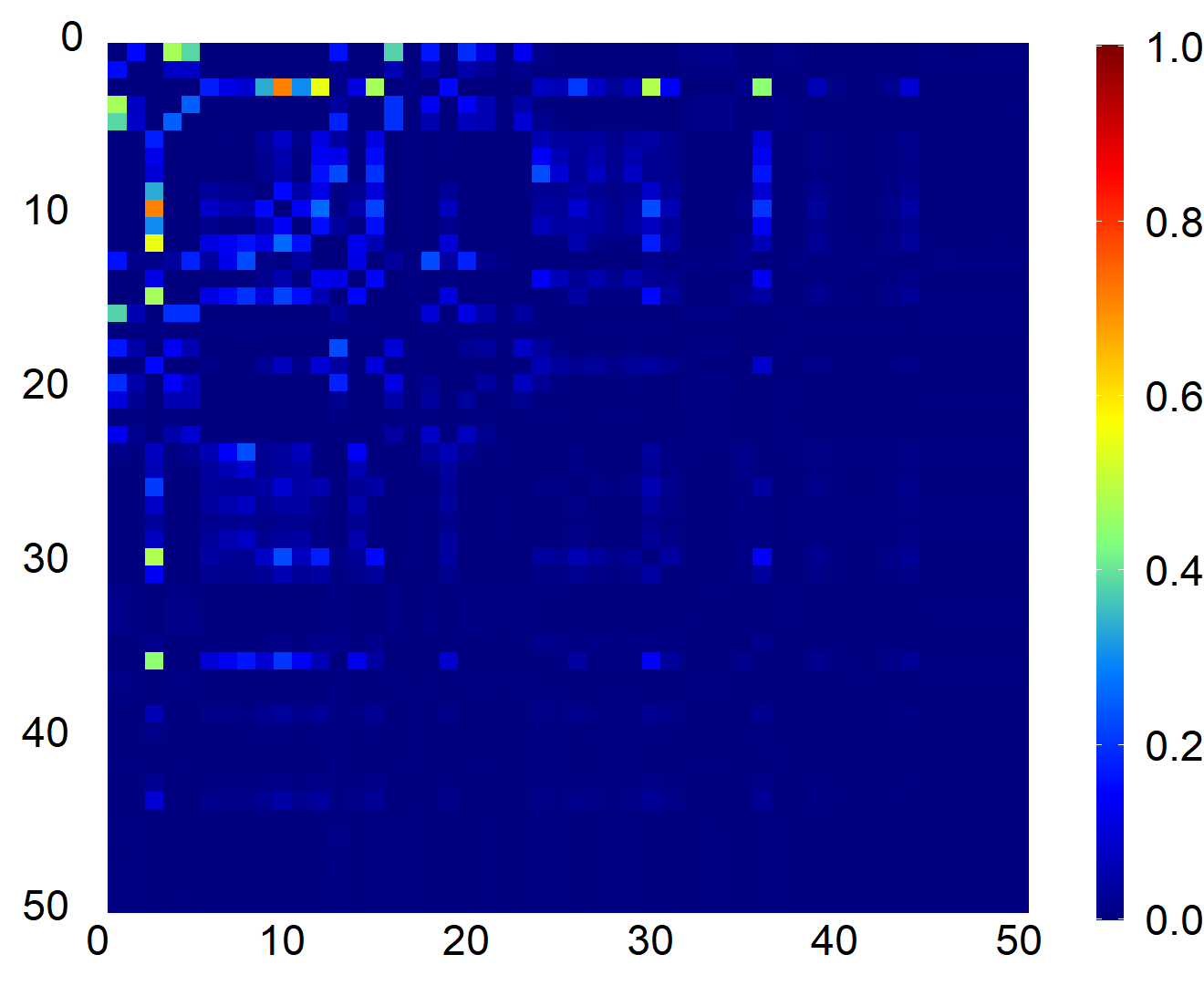} 
		\caption{Group 3 ($12.9\%$ of the subjects)}          
	\end{subfigure}\\
	\begin{subfigure}[t]{0.32\textwidth}
		\centering
		\includegraphics[width=1\linewidth]{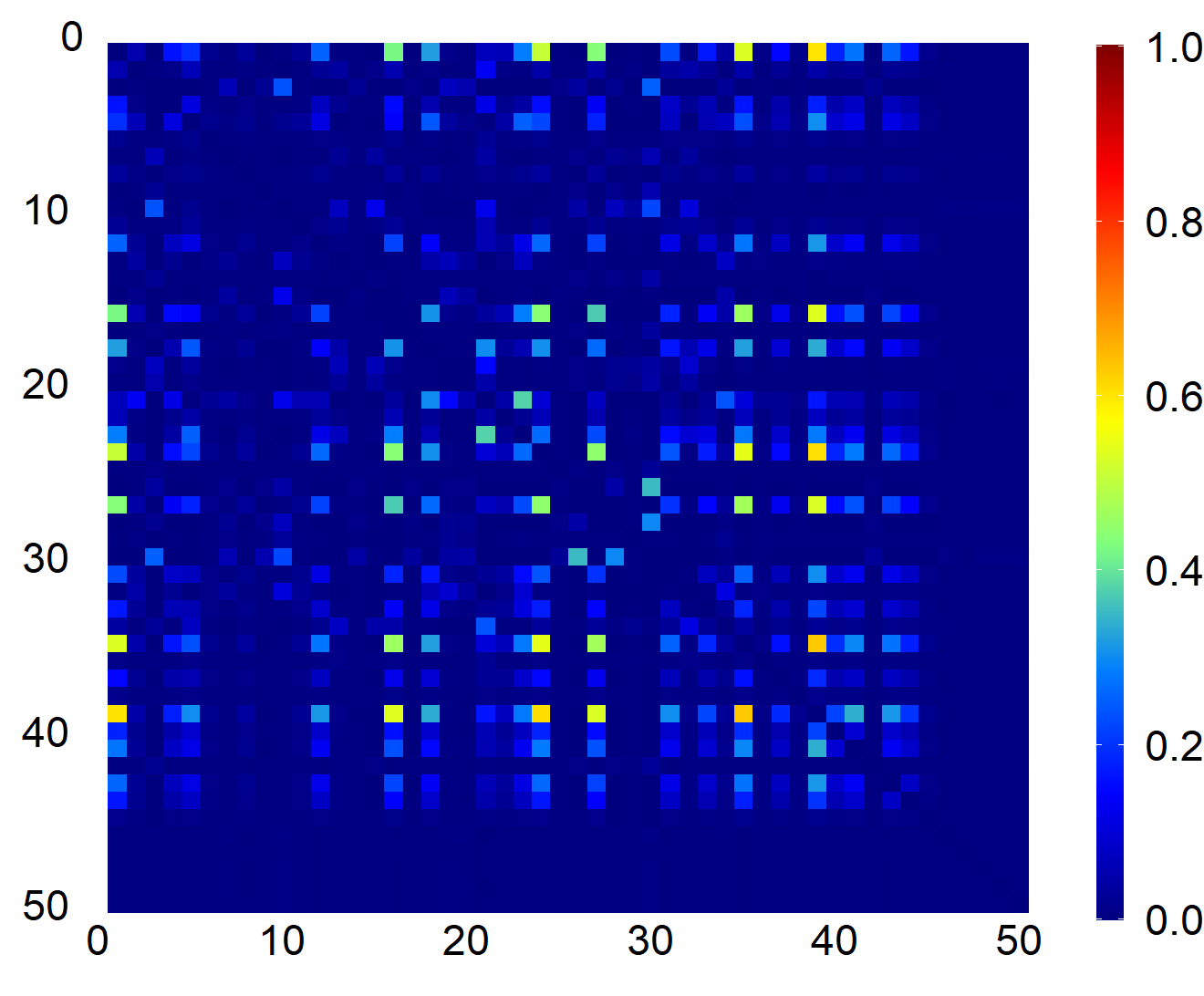}
		\caption{Group 4 ($9.0\%$ of the subjects)}
	\end{subfigure}
	\begin{subfigure}[t]{0.33\textwidth}
		\centering
		\includegraphics[width=0.97\linewidth]{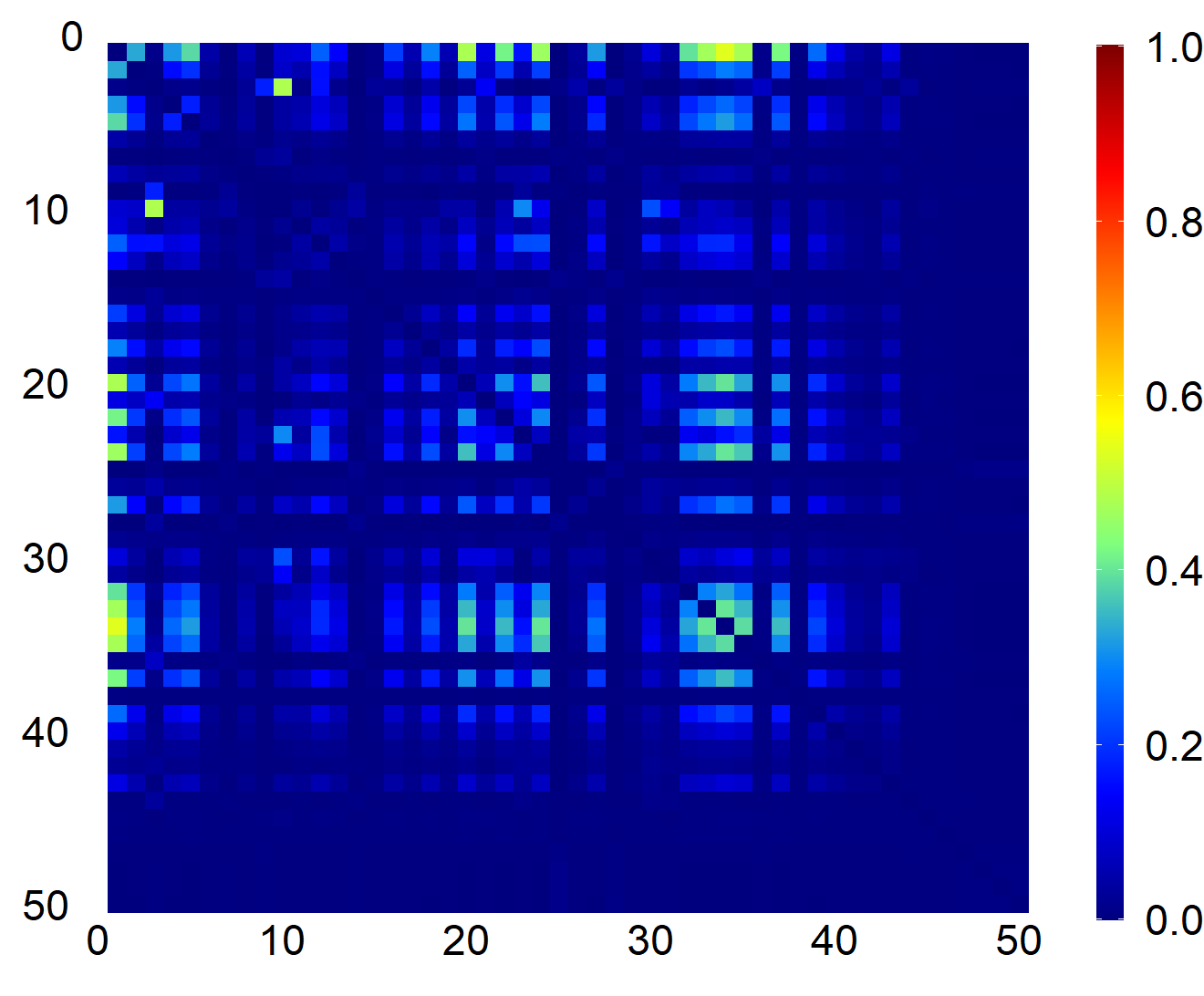}
		\caption{Group 5 ($7.9\%$ of the subjects)}
	\end{subfigure}
	\begin{subfigure}[t]{0.32\textwidth}
		\centering
		\includegraphics[width=1\linewidth]{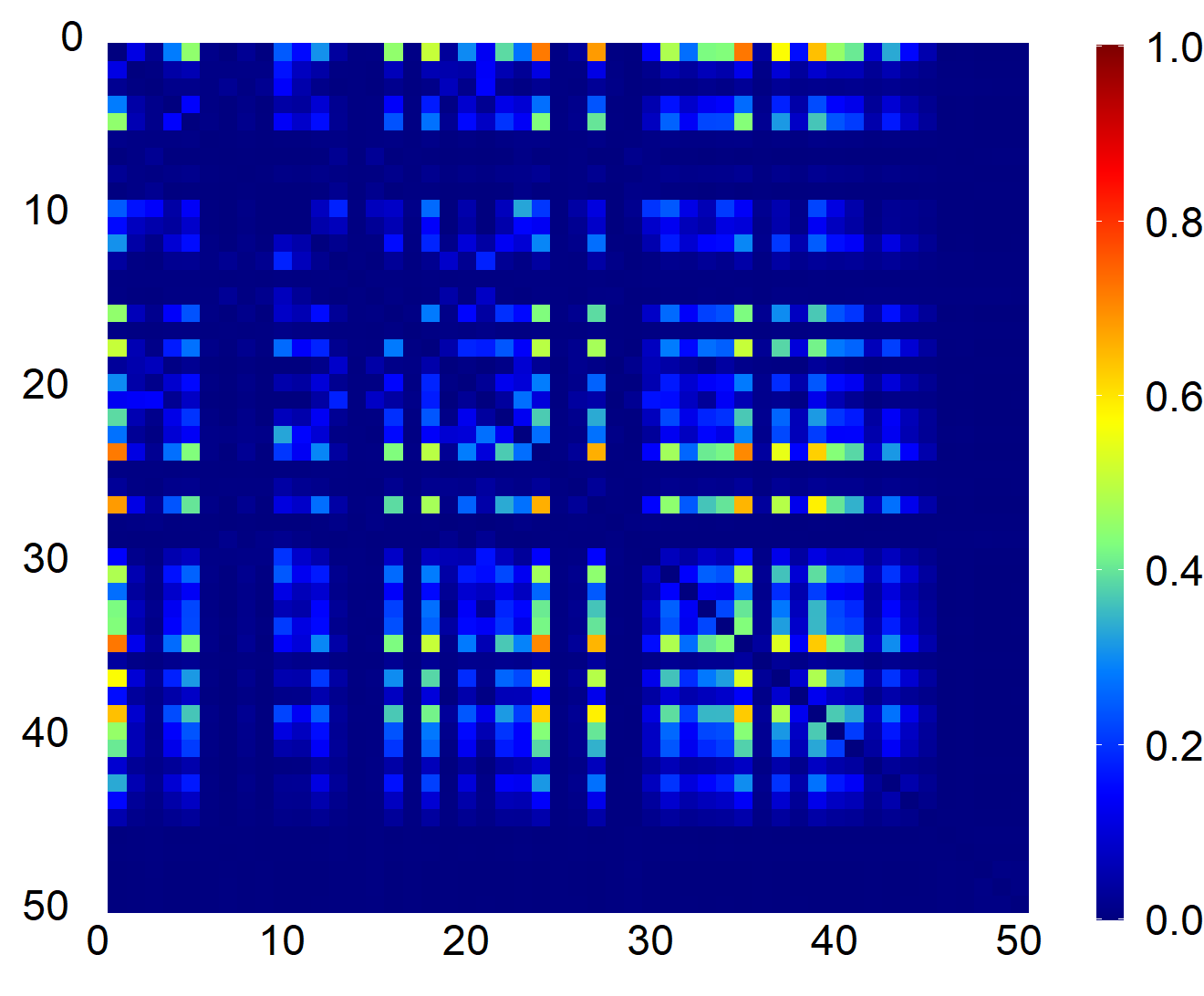}
		\caption{Group 6 ($5.2\%$ of the subjects)}
	\end{subfigure}
	\caption{The posterior means of the edge connectivity probabilities $\Phi(\mu + M_k)$ over the six groups. % (the edges with probability smaller than $0.05$ are hidden for better visualization).  
		\label{fig:fmri2}}
\end{figure}

\cref{fig:fmri} (right) shows the posterior of the number of clusters $T$ for the quasi-Bernoulli mixture model, which is highly concentrated on $T = 6$ clusters. For comparison, we also consider a Dirichlet process mixture and {a Pitman--Yor process mixture.
	We use the same prior for the component parameters $Q_k,\Lambda_k$, as the one in the quasi-Bernoulli model.
	We set the Dirichlet process concentration parameter $\alpha$ so that the expected number of clusters under the Dirichlet process prior is as close as possible to the one under the quasi-Bernoulli process prior for $n=812$.
	Similarly, for the Pitman--Yor process prior ${\text{PY}}(\alpha,d)$, we choose $\alpha$ and $d$ to match both the expectation and the variance of number of clusters with quasi-Bernoulli process prior. The $\alpha$ for Dirichlet process is chosen to be $0.63$, and the parameters of the Pitman--Yor process are $\alpha=0.30,\ d=0.11$.} As the results, these two models yield posterior modes of $T=8$ and $T=7$ clusters (\cref{fig:fmri}, left and middle) and produces several very small clusters ({the Dirichlet process mixture model produces four small groups with $5.8\%$, $5.5\%$, $5.2\%$ and $0.1\%$ of subjects; the Pitman--Yor process mixture model  produces three small groups with $3.3\%$, $2.5\%$ and $1.7\%$ of subjects}). Thus, the result from the quasi-Bernoulli model leads to a more parsimonious representation.

In the quasi-Bernoulli posterior, the subjects are clustered into six groups with high probability. \cref{fig:fmri2} shows the posterior means of the edge probabilities $\Phi(\mu + M_k)$ for each group. These results indicate that sparse connectivity is exhibited by the first three groups of subjects (accounting for $77.8\%$ of subjects), whereas the other three groups have denser connectivity. Specifically, for each group, we examine {the posterior mean proportion} of node pairs with edge connectivity probabilities greater than $0.05$. {The proportions for the six groups are $8.6\%$, $4.4\%$, $8.6\%$, $15.7\%$, $21.1\%$ and $28.2\%$, respectively.}

In addition, we conduct additional experiments using two common Bayesian clustering methods: (i) Dirichlet process mixture of high-dimensional probit-Bernoulli model (without low-rank latent structure); (ii) approximation using mixture of factor analyzers \citep{mclachlan2003modelling} and treating the data as if they were continuous, and selecting the number of clusters using Bayesian information criterion (BIC). The method (i) is a popular solution, however, it is found to suffer from a curse of dimensionality \citep{chandra2020bayesian}. Indeed, applying (i) on the data leads to only $T=1$ cluster in the result. {The method (ii) selects $3$ clusters and $2$ latent dimensions under BIC. The three clusters of the maximum a posteriori (MAP) estimates have $36.6\%$, $20.6\%$ and $42.8\%$ of the subjects. In the appendix, \cref{fig:fmri-mfa} shows the MAP estimate of the mean of each Gaussian component.}
% \vspace{-2cm}

\section{Discussion}
In this article, we propose a modification to the canonical stick-breaking construction, leading to an infinite mixture model that provides consistency for the number of clusters (like the MFM model) as well as easy implementation in posterior MCMC computation (like the Dirichlet process mixture model). \cite{heiner2019structured} similarly proposes a tweak to the breaking proportion $v_k$ under the framework of the Bayesian finite mixture prior, while we focus on the infinite mixture model and construct the theoretical properties on the number of clusters.

There are several extensions worth further pursuing.
First, recovery of the true number of clusters under a {\em misspecified} model is still an open problem. {In a recent work \citep{cai2021finite}, it is theoretically shown that even a small amount of misspecification of the components will tend to result in over-estimation of the number of clusters in overfitted finite mixture models.} Intuitively, this suggests that in addition to controlling the mixture weights to avoid small spurious clusters, it is also important to ensure that the family of component distributions is flexible enough to avoid severe model misspecification issues. Second, it would be interesting to investigate whether the combination of the quasi-Bernoulli infinite mixture framework and distance clustering approaches, such as the Laplacian-based approach \citep{rohe2011spectral}, can lead to a consistency result for the number of clusters.

The popular mixture models (such as Dirichlet process and Pitman--Yor process mixtures) are completely fine for the task of density estimation; nevertheless in some sense, a non-parametric mixing measure entails some misspecification under a indefinitely increasing $n$, which inherently assumes the number of clusters growing with $n$, hence conflicting with the popular modeling view where there is a fixed ground-truth $k_0$. Our insight is that for any fixed $n$, we have an exchangeable partition probability function (which is calculated by integrating out the mixture weights of those non-occupied components) that gives a discrete distribution for $K\in \{1,\ldots,n\}$. Therefore, we can calibrate the asymptotic behavior of the probability function
to produce a consistent estimator, via either controlling $\alpha$ in Dirichlet process mixture or $\epsilon$ as we do in our quasi-Bernoulli model.

%We provide the algorithm in the supplementary materials.
%Given component assignment $c^{(s)}=k$, we have the latent variables 
%\be
%Z^{(s)}_{i,j} \mid c^{(s)}=k &\sim \No\{ (Q_k \Lambda_k Q_k^{\rm T} + D)_{i,j},1 \}\{1(Z^{(s)}_{i,j}>0, A^{(s)}_{i,j} =1) + 1(Z^{(s)}_{i,j}<0, A^{(s)}_{i,j} =0)\},\\
%Z^{(s)}_{i,i} \mid c^{(s)} = k &\sim \No\{ (Q_k \Lambda_k Q_k^{\rm T}+D)_{i,i},2 \},
%\ee
%% which leads to
%% \be
%% \Pi [ Q_k,\Lambda_k \mid \{Z^{(s)}, c^{(s)}\}_{s=1}^n] \propto \exp \biggl\{\ \frac{1}{2}
%% \sum_{s: c^{(s)}=k} \text{tr}( Z^{(s)}  Q_k\Lambda_k Q^{\rm T}_k) - \frac{1}{4} \sum_{s: c^{(s)}=k}\text{tr}(\Lambda_k\Lambda_k) \biggr\}.
%% \ee
%which allows us to update the component parameters from the closed-form full conditional:
%\be
%Q_k &\sim \text{BMF}\biggl(\frac{1}{2}\sum_{s: c^{(s)}=k}\{Z^{(s)}-D\}, \Lambda_k,0\biggr),\\
%\lambda_{k,l} &\sim \No\biggl\{
%\frac{ [Q^{\rm{T}}_{k}\sum_{s: c^{(s)}=k}\{Z^{(s)}-D\}Q_{k}]_{l,l} } {n_k+2/\tau^2},\frac{1}{n_k/2 + 1/\tau^2}
%\biggr \}1(\lambda_{k,l}>0),
%\ee
%where \text{BMF} denotes the matrix Bingham-von Mises-Fisher distribution \citep{hoff2009simulation}.
% \vspace{-2cm}

%\renewcommand{\theHsection}{Asection.\arabic{section}}
%\renewcommand{\theHtable}{Atable.\arabic{table}}
%\renewcommand{\theHfigure}{Afigure.\arabic{figure}}
%\renewcommand{\theHequation}{Aequation.\arabic{section}.\arabic{equation}}

\appendix

\section{Proofs}
\label{sec:proofs}

\begin{proof}{\bf of \cref{thm:eppf}}~\\
	The conditional probability mass function of the assignment variables $c = (c_1,\ldots,c_n)$ is
	\[
	\Pr(c \mid b_1,b_2,\ldots,\beta_1,\beta_2,\ldots) = \prod_{k=1}^{\infty} (1-\beta_k b_k)^{n_k} (\beta_k b_k)^{m_k},
	\]
	where $n_k=\sum_{i=1}^n \mathds{1}(c_i=k)$
	and $m_k=\sum_{i=1}^n \mathds{1}(c_i>k)$.
	Define $M(c) := \max\{c_1,\dots,c_n\}$ and $Q_k := \tilde{p} + (1-\tilde{p}) I_\epsilon(m_k+\alpha, n_k+1) / \epsilon^\alpha$. Then
	\be
	\Pr({c}) &= \prod_{k=1}^{\infty} \int^1_{{0}}\! \Big(\tilde{p}   (1-\beta_k )^{n_k} (\beta_k )^{m_k}+(1-\tilde{p}) (1-\epsilon\beta_k)^{n_k}(\epsilon\beta_k)^{m_k}\Big)\alpha\beta_{k}^{\alpha-1}\,d\beta_{k}\\
	&=  \prod_{k=1}^{\infty} \Big( \tilde{p}\alpha \mathrm{B}(n_k+1,  m_k + \alpha) + (1-\tilde{p})\frac{\alpha}{\epsilon^\alpha}\mathrm{B}(n_k+1,m_k+\alpha)\int_{0}^{\epsilon}\frac{(1-x)^{n_k}(x)^{m_k+\alpha-1}}{\mathrm{B}(n_k+1,m_k+\alpha)}\,d x  \Big)  \\ 
	&\stackrel{\textup{(a)}}{=} \prod_{k=1}^{M({c} )} \Big( \tilde{p}\alpha \mathrm{B}(n_k+1,  m_k + \alpha)  + (1-\tilde{p})\frac{\alpha}{\epsilon^\alpha}\mathrm{B}(n_k+1,m_k+\alpha)I_{\epsilon}(m_k+\alpha, n_k+1)  \Big)\\
	&= \prod_{k=1}^{M(c)} \frac{\alpha\Gamma(n_k+1)\Gamma(m_k+\alpha)}{\Gamma(n_k+m_k+\alpha+1)} Q_k \\
	&\stackrel{\textup{(b)}}{=} \bigg( \prod_{k=1}^{M(c)} \Gamma(n_k+1) \bigg) \prod_{k=1}^{M(c)} \frac{\alpha Q_k \Gamma(m_k+\alpha)}{(m_{k-1} + \alpha) \Gamma(m_{k-1} + \alpha)} \\
	%&= \prod_{k=1}^{M(c)} \Gamma(n_k+1) \prod_{k=1}^{M(c)} \frac{\Gamma(m_k + \alpha)} {\Gamma(g_k+\alpha+1)} \prod_{k=1}^{M(c)} \alpha \Big( \tilde{p} +(1-\tilde{p})\frac{1}{\epsilon^\alpha}I_{\epsilon}(m_k+\alpha, n_k+1)\Big)\\
	&= \frac{\Gamma(\alpha)}{\Gamma(n+\alpha)} \bigg( \prod_{k=1}^{M(c)} \Gamma(n_k+1) \bigg) \prod_{k=1}^{M(c)} \frac{\alpha Q_k}{m_{k-1}+\alpha}.
	\ee
	In step (a), we use the fact that for all $k > M(c)$, we have $n_k=0$ and $m_k=0$,
	and thus, everything cancels for such $k$ since $B(1,\alpha)=1/\alpha$ and $I_{\epsilon}(\alpha, 1)=\epsilon^\alpha$.
	In step (b), we use the fact that $n_k + m_k = m_{k-1}$, and thus, $\Gamma(n_k + m_k + \alpha + 1) = (m_{k-1} + \alpha) \Gamma(m_{k-1} + \alpha)$.
	
	Define $g_{k} := \sum_{i=1}^{n} \mathds{1}(c_i\geq k) = \sum_{l=k}^{\infty} n_l$, and note that $g_{k} = m_{k-1}$.
	Let $\mathcal{A}_{{c}}$ be the partition of $\{1,\dots,n\}$ induced by $c$. Fix a partition $\mathcal{A}=\{A_1,\dots,A_t\}$.
	When $\mathcal{A}_{{c}}=\mathcal{A}$, there are exactly $t$ unique values among $c_1,\dots,c_n$. Let $k_1<k_2<\cdots<k_t$ denote these unique values, and set $k_0=0$.
	For $k_{j-1}<k<k_{j}$, we have $n_k=0$ and $g_{k} = g_{k+1} = g_{k_j}$, and thus
	$ Q_k = \tilde{p} + (1-\tilde{p}) \epsilon^{g_{k_j}}$. Meanwhile, for $k=k_{j}$, we have $n_k=n_{k_j}$ and $g_{k+1} = g_{k_{j+1}}$, and thus it follows that
	$ Q_k = \tilde{p} + (1-\tilde{p}) I_\epsilon(g_{k_{j+1}}+\alpha, n_{k_j}+1) / \epsilon^\alpha $. Hence, for all ${c}$ such that $\mathcal{A}_{{c}}=\mathcal{A}$, we have
	\[
	\Pr({c}) = \frac{\Gamma(\alpha)}{\Gamma(n+\alpha)} \bigg( \prod_{j=1}^t \Gamma(n_{k_j}+1) \bigg) \prod_{j=1}^{t}  U_j({c})
	\]
	where
	\[
	U_j({c}) = \bigg(
	\frac {\alpha \tilde{p} + \alpha(1-\tilde{p}) I_\epsilon(g_{k_{j+1}}+\alpha, n_{k_j}+1)/\epsilon^\alpha}{ g_{k_j}+\alpha}
	\bigg)
	\bigg( \frac{\alpha \tilde{p} + \alpha (1-\tilde{p}) \epsilon^{g_{k_j}}}{g_{k_j}+\alpha} \bigg)^{d_j}
	\]
	where $d_j = k_j-k_{j-1}-1$.
	Since there is a unique permutation ${\sigma}=(\sigma_1,\dots,\sigma_t)$ of $\{1,\dots,t\}$ such that $A_{\sigma_j}=\{i:c_i=k_j\}$ for all $j\in\{1,\ldots,t\}$, the mapping between $\{{c}:\mathcal{A}_{{c}}=\mathcal{A}\}$ and $\{ ({\sigma}, d_1,\dots,d_t) : {\sigma} \in S_t,\ d_1,\ldots,d_t\in\mathbb{N}\}$ is a bijection.
	(Here, $S_t$ is the set of all permutations of $\{1,\dots,t\}$, and $\mathbb{N} := \{0,1,2,\ldots\}$.)
	Let $n^*_j := |A_j|$ and $g_j^*(\sigma) := \sum_{l=j}^t n^*_{\sigma_l}$.
	For the value of $(\sigma,d_{1:t})$ that corresponds to $c$, we have $g_{k_j} = g^*_j(\sigma)$ and $n_{k_j} = |A_{\sigma_j}| = n^*_{\sigma_j}$, and thus, $U_j(c) = U^*_j(\sigma,d_{1:t})$ where
	\[
	U^*_j(\sigma,d_{1:t})
	:= \bigg(
	\frac {\alpha \tilde{p} + \alpha(1-\tilde{p}) I_\epsilon(g^*_{j+1}(\sigma)+\alpha, n^*_{\sigma_j}+1)/\epsilon^\alpha}{ g^*_j(\sigma)+\alpha }
	\bigg)
	\bigg( \frac{\alpha \tilde{p} + \alpha (1-\tilde{p}) \epsilon^{g^*_j(\sigma)}}{g^*_j(\sigma)+\alpha} \bigg)^{d_j}.
	\]
	Summing over $d_j$ and using $\sum_{d=0}^\infty x^d = 1/(1-x)$ for $|x|<1$, we have
	\be
	\sum_{d_j=0}^\infty U^*_j(\sigma,d_{1:t})
	&= \frac{\alpha \tilde{p} + \alpha(1-\tilde{p}) I_\epsilon(g^*_{j+1}(\sigma)+\alpha, n^*_{\sigma_j}+1)/\epsilon^\alpha}{g^*_j(\sigma)+\alpha(1 - \tilde{p})(1 - \epsilon^{g^*_j(\sigma)})}.
	\ee
	Note that $U_j^*(\sigma,d_{1:t})$ depends on $d_{1:t}$ only through $d_j$.
	Therefore,
	\be
	\Pr_{\epsilon,n}(\mathcal{A}) &= \sum_{c\,:\,\mathcal{A}_{{c}}=\mathcal{A}}\Pr({c}) = \frac {\Gamma(\alpha)} {\Gamma(n+\alpha)} \bigg(\prod_{j=1}^t \Gamma(n^*_j+1)\bigg)
	\sum_{{\sigma} \in S_t} \sum_{d_1=0}^\infty \cdots \sum_{d_t=0}^\infty \prod_{j=1}^{t}  U^*_j(\sigma,d_{1:t}) \\
	&= \frac {\Gamma(\alpha)} {\Gamma(n+\alpha)} \bigg(\prod_{j=1}^t \Gamma(n^*_j+1)\bigg)
	\sum_{\text{all }{\sigma}}\prod_{j=1}^{t}   \sum_{d_j=0}^\infty U^*_j(\sigma,d_{1:t}) \\
	&= \frac {\alpha^t\Gamma(\alpha)} {\Gamma(n+\alpha)} \bigg(\prod_{j=1}^t \Gamma(n^*_j+1)\bigg)
	\sum_{\text{all }{\sigma}}\prod_{j=1}^{t} \frac{\tilde{p} + (1-\tilde{p}) I_\epsilon(g^*_{j+1}(\sigma)+\alpha, n^*_{\sigma_j}+1)/\epsilon^\alpha}{g^*_j(\sigma)+\alpha(1 - \tilde{p})(1 - \epsilon^{g^*_j(\sigma)})}.
	\ee
	This proves the result.
\end{proof}

\vspace{1em}
\begin{proof}{\bf of Lemma~\ref{lemma:eppf0}}~\\
	The general approach of the proof follows the technique of \citet{miller2019elementary}.
	The conditional probability mass function of the assignment variables is
	\[
	\Pr({c} \mid b_1,b_2,\ldots,\beta_1,\beta_2,\ldots) = \prod_{k=1}^{\infty} (1-\beta_k b_k)^{n_k} (\beta_k b_k)^{g_{k+1}},
	\]
	where $n_k=\sum_{i=1}^n \mathds{1}(c_i=k)$ and $g_{k}=\sum_{i=1}^{n} \mathds{1}(c_i\geq k)$. Let $M({c} )= \max\{c_1,\dots,c_n\}$. Then
	\begin{align*} 
		\Pr({c}) &= \prod_{k=1}^{\infty} \int^1_{{0}}\! \Big(\tilde{p}   (1-\beta_k )^{n_k} (\beta_k )^{g_{k+1}}+(1-\tilde{p})\mathds{1}(g_{k+1}=0) \Big)\alpha\beta_{k}^{\alpha-1}\,d\beta_{k}\\
		&=\prod_{k=1}^{M({c} )} \Big( \tilde{p}\alpha \mathrm{B}(n_k+1,  g_{k+1} + \alpha)  + (1-\tilde{p}) \mathds{1}(g_{k+1}=0) \Big)\\
		&= \bigg(\prod_{k=1}^{M(c)-1} \frac{\tilde{p}\alpha\Gamma(n_k+1)\Gamma(g_{k+1}+\alpha)}{\Gamma(n_k+g_{k+1}+\alpha+1)} \bigg) \big(\tilde{p}\alpha\mathrm{B}(n_{M(c)}+1,\alpha)+1-\tilde{p}\big) \\
		&= \bigg(\prod_{k=1}^{M(c)-1} \Gamma(n_k+1)\bigg) \bigg(\prod_{k=1}^{M(c)-1} \frac{\tilde{p}\alpha\Gamma(g_{k+1} + \alpha)} {\Gamma(g_k+\alpha+1)}\bigg) \big( \tilde{p}\alpha\mathrm{B}(n_{M(c)}+1,\alpha)+1-\tilde{p}\big)\\
		&= \frac{\Gamma(\alpha)}{\Gamma(n+\alpha)} \bigg( \prod_{k=1}^{M(c)} \Gamma(n_k+1) \bigg) \bigg(\prod_{k=1}^{M(c)-1} \frac{\tilde{p}\alpha }{g_{k}+\alpha}\bigg) \bigg( \frac{\tilde{p}\alpha + (1-\tilde{p})/\mathrm{B}(n_{M(c)}+1,\alpha)}{n_{M(c)}+\alpha}\bigg).
	\end{align*}
	
	Let $\mathcal{A}({c})$ denote the partition of $\{1,\dots,n\}$ corresponding to ${c}$. For fixed $\mathcal{A}=\{A_1,\dots,A_t\}$, when $\mathcal{A}({c})=\mathcal{A}$, there are exactly $t$ unique values among $c_1,\dots,c_n$. Let $k_1<k_2<\cdots<k_t$ denote these unique values, and set $k_0=0$.
	For $k_{j-1}<k \leq k_{j}$, we have $g_{k} = g_{k_j}$. Hence, for ${c}$ satisfying $\mathcal{A}({c})=\mathcal{A}$, we have
	$
	\Pr({c}) = (\Gamma(\alpha)/\Gamma(n+\alpha)) \big( \prod_{j=1}^t \Gamma(n_{k_j}+1) \big) \prod_{j=1}^{t}  U_j({c})
	$, where $U_j({c}) = (\tilde{p}\alpha / (g_{k_j}+\alpha))^{d_j}$ for $1\leq j<t$ and
	\[
	U_t({c}) = \frac{\tilde{p}\alpha + (1-\tilde{p})/\mathrm{B}(n_{k_t}+1,\alpha)}{n_{k_t}+\alpha} 
	\biggl( \frac{\tilde{p} \alpha}{g_{k_t}+\alpha} \biggr) ^ {d_t-1}
	= \bigg(1 + \frac{(1-\tilde{p})/\tilde{p} \alpha}{\mathrm{B}(n_{k_t}+1,\alpha)}\bigg)\bigg(\frac{\tilde{p}\alpha}{g_{k_t}+1}\bigg)^{d_t}
	\]
	where $d_j = k_j-k_{j-1}$ for $j=1,\dots,t$.

	Since there is a unique permutation ${\sigma}=(\sigma_1,\dots,\sigma_t)\in S_t$  such that $A_{\sigma_j}=\{i:c_i=k_j\}$, the mapping between $\{{c}:\mathcal{A}({c})=\mathcal{A}\}$ and $ \big\{ ({\sigma}, d_1,\dots,d_t) : {\sigma} \in S_t, \ d_1,\dots,d_t\in\{1,2,\ldots\} \big\} $ is a bijection.
	Letting $n_j=|A_j|$, we have
	\begin{align}\label{eq:p_A}
		\Pr_{0,n}(\mathcal{A}) = \sum_{c\,:\,\mathcal{A}({c})=\mathcal{A}}\Pr({c}) = \frac {\Gamma(\alpha)} {\Gamma(n+\alpha)} \bigg(\prod_{j=1}^t \Gamma(n_{j}+1)\bigg)
		\sum_{{\sigma}\in S_t}\sum_{d_1=1}^\infty \cdots \sum_{d_t=1}^\infty \prod_{j=1}^{t}  U_j({c}),
	\end{align}
	treating $c$ as a function of $\sigma,d_1,\ldots,d_t$.
	Changing the order of summations and multiplication, defining $g_j(\sigma)=\sum_{l=j}^t n_{\sigma_l}$, and using the geometric series $\sum_{d=1}^\infty x^d = x/(1-x)$ for $x\in (0,1)$,
	\[
	\begin{aligned}
		\sum_{d_1=1}^\infty \cdots \sum_{d_t=1}^\infty \prod_{j=1}^{t}  U_j({c})
		&= \bigg(1 + \frac{(1-\tilde{p})/\tilde{p} \alpha}{\mathrm{B}(n_{\sigma_t}+1,\alpha)}\bigg)
		\prod_{j=1}^{t} 
		\sum_{d_j=1}^{\infty} \biggl( \frac{ \tilde{p} \alpha } {g_j(\sigma)+\alpha} \biggr)^{d_j} \\
		&= \bigg(1 + \frac{(1-\tilde{p})/\tilde{p} \alpha}{\mathrm{B}(n_{\sigma_t}+1,\alpha)}\bigg)
		\prod_{j=1}^{t} \frac{ \tilde{p} \alpha } {g_j(\sigma)+\alpha(1-\tilde{p})} \\
		&= \alpha^t \prod_{j=1}^{t} \frac{ \tilde{p} + \mathds{1}(j=t)(1-\tilde{p})/(\alpha \mathrm{B}(n_{\sigma_t}+1,\alpha))} {g_j(\sigma)+\alpha(1-\tilde{p})}.
	\end{aligned}
	\]
	Combining with \cref{eq:p_A}, this proves the result.
\end{proof}

	\vspace{1em}
	We provide the complete statements of the two results from \cite{nobile1994bayesian}.
	\begin{theorem}{\citet[Corollary 3.1]{nobile1994bayesian}}\label{thm:nobile1}
		Assume $\phi\in \Omega'$ is identifiable up to permutation of the mixture components.
		Let $\Pi_0$ be the prior on $\Omega$ under the model defined by \cref{eq:qb0,eq:clusters}, and assume $\Pi_0(\{\phi:\exists\ i\neq j \text{ such that } \theta_i=\theta_j\})=0$. Let $\Pi_0'$ be the corresponding prior on $\Omega'$ induced by $\eta$.
		Then there is a subset $\Omega'_0 \subset \Omega'$ with $\Pi'_0(\Omega'_0) = 1$ such that for any $\phi_0 = (k_0,w^0_1,\ldots,w^0_{k_0},\theta^0_1,\ldots,\theta^0_{k_0}) \in \Omega'_0$,
		if $y_1,y_2,\ldots \mid \phi_0 \stackrel{iid}\sim P_{\phi_0}$, then as $n\to\infty$, we have
		\be
		\Pr_{\epsilon=0}(\phi \in D \mid y_{1:n}) &\to \mathds{1}(\phi_0\in D') \quad \mathrm{a.s.}[P_{\phi_0}],
		\ee
		for any measurable subset $D'\subset \Omega'$ and $D=\{\phi \in \Omega: \eta(\phi)\in D'\}$.
	\end{theorem}

\vspace{1em}
	
	\begin{theorem}{\citet[Proposition~3.5]{nobile1994bayesian}}\label{thm:nobile2}
		Under the same assumptions of \cref{thm:nobile1},
		\be
		\Pr_{\epsilon=0}(K=k \mid y_{1:n}) &\to \mathds{1}(k_0=k) \quad \mathrm{a.s.}[P_{\phi_0}].
		\ee
	\end{theorem}

\vspace{1em}
\begin{proof}{\bf of \cref{thm:consistent0}}~\\
	The first result is proved by \cref{thm:nobile2}. The second result can be proved as following. Using the \cref{thm:nobile1}, there is a subspace $\Omega'_0$ as described in the theorem such that if $\phi_0\in \Omega'_0$ then the posterior distribution of $(w,\theta)$ given $K=k_0$ and $y_{1:n}$ will converge to {a uniform distribution in which with equal probability the $(w,\theta)$ is one of the permutation of $(w^0_1,\theta^0_1),\dots,(w^0_{k_0},\theta^0_{k_0})$. This is because the transformation $\eta$ maps all of the permutations of $(w^0_1,\theta^0_1),\dots,(w^0_{k_0},\theta^0_{k_0})$ into the same one with a specific order.
	}
	Define the random variables $N_k = \sum_{i=1}^n \mathds{1}(c_i=k)$ for $k=1,\dots,k_0$. Then
	\begin{align}
		& p_{\epsilon=0}(N_k=0 \mid K=k_0, y_{1:n}) \notag\\
		& = \Pr_{\epsilon = 0}(\cap_{i=1}^n\{c_i\neq k\} \mid K=k_0, y_{1:n}) \notag\\
		& = \int \Pr_{\epsilon=0} ( \cap_{i=1}^n\{c_i\neq k\} \mid K=k_0, w,\theta,y_{1:n}) \mathbb{P}_{\epsilon=0} (dw,d\theta \mid K=k_0,y_{1:n}) \notag\\
		& = \int \prod_{i=1}^n \biggl( 1 - \frac{w_k f_{\theta_k}(y_i)}{\sum_{l=1}^{k_0} w_l f_{\theta_l}(y_i)} \biggr) \mathbb{P}_{\epsilon=0} (dw, d\theta \mid K=k_0,y_{1:n})\notag\\
		& \leq \int \prod_{i=1}^{n_0} \biggl( 1 - \frac{w_k f_{\theta_k}(y_i)}{\sum_{l=1}^{k_0} w_l f_{\theta_l}(y_i)} \biggr) \mathbb{P}_{\epsilon=0} (dw, d\theta \mid K=k_0,y_{1:n})\label{eq:pfthm3}
	\end{align}
	{for any given positive integer $n_0$ and $n\geq n_0$. Using the weak convergence of the posterior distribution of the $(w,\theta)$ and the fact that the integrand is bounded and $f_{\theta}$ is continuous at all $\theta^0_{k}$'s, the 
		\cref{eq:pfthm3} converges to 
		\[
		\sum_{k=1}^{k_0} \frac{1}{k_0} \prod_{i=1}^{n_0} \biggl( 1 - \frac{w^0_{k} f_{\theta^0_{k}}(y_i)}{\sum_{l=1}^{k_0} w^0_l f_{\theta^0_l}(y_i)} \biggr)
		\]
		$\mathrm{a.s.}[P_{\phi_0}]$ when $n\to\infty$. Since \cref{eq:pfthm3} holds for any positive integer $n_0$, we have
		\[
		\limsup_{n\to\infty} p_{\epsilon=0}(N_k=0 \mid K=k_0, y_{1:n}) \leq \sum_{k=1}^{k_0} \frac{1}{k_0} \prod_{i=1}^{\infty} \biggl( 1 - \frac{w^0_{k} f_{\theta^0_{k}}(y_i)}{\sum_{l=1}^{k_0} w^0_l f_{\theta^0_l}(y_i)} \biggr).
		\]
	}
	For every $k=1,\dots,k_0$, there exists a measurable set $D_k$ with non-zero measure such that $f_{\theta^0_k}(y) \geq \delta_k >0$ when $y\in D_k$, and all $f_{\theta^0_l}(y)\ (l=1,\dots,k_0)$ are finite on $D_k$.
	For every $k=1,\dots,k_0$, there exists a sequence $n_{k1},n_{k2},\dots$ such that $y_{n_{ki}} \in D_k\ (i=1,2,\dots)$, $\mathrm{a.s.}[P_{\phi_0}]$. Hence, for all $i\geq 1$, $f_{\theta^0_k}(y_{n_{ki}}) \geq \delta_k$ and $\sum_{l=1}^{k_0} w^0_l f_{\theta^0_l}(y_{n_{ki}}) \leq M$ for some $M$. Therefore,
	\[
	\prod_{i=1}^\infty \biggl( 1 - \frac{w^0_{k} f_{\theta^0_{k}}(y_i)}{\sum_{l=1}^{k_0} w^0_l f_{\theta^0_l}(y_i)} \biggr) 
	\leq \prod_{i=1}^\infty \biggl( 1 - \frac{w^0_{k} f_{\theta^0_{k}}(y_{n_{ki}})}{\sum_{l=1}^{k_0} w^0_l f_{\theta^0_l}(y_{n_{ki}})} \biggr)
	\leq \prod_{i=1}^\infty \biggl( 1 - \frac{w^0_{k} \delta}{M} \biggr) = 0,
	\]
	which leads to $p_{\epsilon=0}(N_k=0 \mid K=k_0, y_{1:n}) \stackrel{\mathrm{a.s.}}{\longrightarrow} 0$ as $n\to\infty$.
	Hence, given $K=k_0$, the posterior probability of having $k_0$ clusters is
	\be
	& \Pr_{\epsilon=0} ( T = k_0 \mid K = k_0,y_{1:n}  )  
	= \Pr_{\epsilon=0} ( N_1>0,\dots,N_{k_0}>0 \mid K=k_0, y_{1:n})\\
	& = 1 - \Pr_{\epsilon=0} ( \cup_{k=1}^{k_0} \{N_k=0\} \mid K=k_0, y_{1:n})\\
	& \geq 1 - \sum_{k=1}^{k_0} \Pr_{\epsilon=0} ( N_k=0 \mid K=k_0, y_{1:n}) \to 1
	\ee
	$\mathrm{a.s.}[P_{\phi_0}]$ as $n\to\infty$. Therefore,
	\be
	\Pr_{\epsilon=0}(T=k_0 \mid y_{1:n}) 
	& = \sum_{k=k_0}^\infty \Pr_{\epsilon=0}(T=k_0 \mid K=k,y_{1:n})\Pr_{\epsilon=0}(K=k \mid y_{1:n}) \\
	& \geq \Pr_{\epsilon=0}(T=k_0 \mid K=k_0, y_{1:n})\Pr_{\epsilon=0}(K=k_0 \mid y_{1:n}) \to 1
	\ee
	$\mathrm{a.s.}[P_{\phi_0}]$ as $n\to\infty$.
\end{proof}

\vspace{1em}
\begin{proof}{\bf of \cref{thm:prior}}~\\
	For a given partition $\mathcal{A}$, let $t=|\mathcal{A}|$.   Define the following notation to represent the factors in  $\Pr_{0,n}(\mathcal{A})$ and $\Pr_{\epsilon,n}(\mathcal{A})$, respectively:
	\be
	U_j({\sigma}) &:= \frac{\tilde{p} + \mathds{1}(j=t)(1-\tilde{p}) / (\alpha \mathrm{B}(\alpha, n_{\sigma_t}+1))}
	{ g_j(\sigma) + \alpha (1 - \tilde{p}) }\\
	V_j({\sigma}) &:= \frac{\tilde{p} + (1-\tilde{p}) I_{\epsilon}( g_{j+1}(\sigma) + \alpha, n_{\sigma_j}+ 1 )/{\epsilon^\alpha} } 
	{ g_j(\sigma) + \alpha (1 - \tilde{p}) (1 - \epsilon^{g_j(\sigma)}) }
	\ee
	for $j=1,\ldots,t$.
	When $j<t$, we have
	\bel\label{eq:AB1}
	U_j(\sigma) \leq  V_j(\sigma)
	\eel
	since $(1-\tilde{p}) I_{\epsilon}( g_{j+1}(\sigma) + \alpha, n_{\sigma_j}+ 1 )/{\epsilon^\alpha} > 0$ and $g_j(\sigma) + \alpha (1 - \tilde{p})>g_j(\sigma) + \alpha (1 - \tilde{p}) (1 - \epsilon^{g_j(\sigma)})>0$.
	
	Meanwhile, for the case of $j=t$, we have
	\bel\label{eq:AB2}
	\frac{U_t({\sigma})}{V_t({\sigma})} 
	& \stackrel{\textup{(a)}}{\leq} \frac{\tilde{p} + (1-\tilde{p}) / ( \alpha \mathrm{B} (\alpha, n_{\sigma_t}+1) ) }{\tilde{p} + (1-\tilde{p}) I_{\epsilon}( \alpha, n_{\sigma_t}+ 1 )/{\epsilon^\alpha}} \\
	& \stackrel{\textup{(b)}}{\leq} \frac{\tilde{p}\alpha \mathrm{B} (\alpha, n_{\sigma_t}+1)+1-\tilde{p}}{\tilde{p}\alpha \mathrm{B} (\alpha, n_{\sigma_t}+1) + (1-\tilde{p})(1-\alpha\epsilon n_{\sigma_t}/(\alpha+1))}\\
	& = 1 + \frac{(1-\tilde{p})\alpha\epsilon n_{\sigma_t}/(\alpha+1)}{\tilde{p}\alpha \mathrm{B} (\alpha, n_{\sigma_t}+1) + (1-\tilde{p})(1-\alpha\epsilon n_{\sigma_t}/(\alpha+1))}\\
	& \stackrel{\textup{(c)}}{\leq} 1 + \frac{\alpha n\epsilon}{\alpha + 1-\alpha\epsilon n } ,
	\eel
	where (a) uses $g_{\sigma_k} + \alpha (1 - \tilde{p})>g_{\sigma_k} + \alpha (1 - \tilde{p}) (1 - \epsilon^{g_{\sigma_k}})>0$, (b) uses
	\be
	&\frac{\alpha \mathrm{B} (\alpha, n_{\sigma_t}+1) I_{\epsilon}( \alpha, n_{\sigma_t}+ 1 )}{\epsilon^\alpha}
	= \frac{\alpha}{\epsilon^\alpha}\int_{0}^{\epsilon} x^{\alpha-1}(1-x)^{n_{\sigma_t}}\,d x \geq \frac{\alpha}{\epsilon^\alpha}\int_{0}^{\epsilon} x^{\alpha-1}(1-n_{\sigma_t} x)\,d x
	\\ &= 1-\frac{\alpha\epsilon n_{\sigma_t}}{\alpha+1},
	\ee
	and (c) uses $\tilde{p}\alpha \mathrm{B} (\alpha, n_{\sigma_t}+1)>0$, $n_{\sigma_t}\leq n$, and the assumption that $\epsilon \leq 1/n$. 
	
	Using the exchangeable partition probability functions in \cref{thm:eppf} and Lemma \ref{lemma:eppf0}, along with \cref{eq:AB1,eq:AB2,eq:ratio_of_sums_inequality}, we have
	\bel\label{eq:pmfratio}
	& \frac{\Pr_{0,n}(\mathcal{A})}{\Pr_{\epsilon,n}(\mathcal{A})} 
	=
	\frac{\sum_{{\sigma} \in S_t }  \prod_{j=1}^{t} U_j({\sigma})}
	{ \sum_{{\sigma}\in S_t}  \prod_{j=1}^{t}V_j({\sigma})  }  \leq  1 + \frac{\alpha n\epsilon}{\alpha+1-\alpha\epsilon n}
	\eel
	for all partitions $\mathcal{A}$.
	Therefore, the Kullback--Leibler divergence satisfies
	\[
	D_{\text{KL}}(\Pr_{0,n} \| \Pr_{\epsilon,n}) 
	= \sum_{\mathcal{A}\in \cup_{t=1}^n \mathcal{H}_t(n)} \Pr_{0,n}(\mathcal{A}) \log \frac{\Pr_{0,n}(\mathcal{A})}{\Pr_{\epsilon,n}(\mathcal{A})} 
	\leq \sum_{\mathcal{A}} \Pr_{0,n}(\mathcal{A}) \frac{\alpha n\epsilon}{\alpha+1-\alpha\epsilon n}
	=  \frac{\alpha n\epsilon}{\alpha+1-\alpha\epsilon n} ,
	\]
	where the sum is over all partitions of $\{1,\dots,n\}$ and the inequality uses $\log(x)\leq x-1$. The result follows by Pinsker's inequality.
\end{proof}

\vspace{1em}
In the proof of \cref{thm:consistent}, we employ the following two inequalities. Let $x_i,y_i\geq 0$ for all $i$ in some countable set $I$, such that $\sum_{i\in I} x_i > 0$ and $\sum_{i\in I} y_i > 0$.
First, if $A \subseteq I$ then
\bel\label{eq:difference_of_ratios_inequality}
\bigg| \frac{\sum_{i\in A} x_i}{\sum_{i\in I} x_i} - \frac{\sum_{i\in A} y_i}{\sum_{i\in I} y_i}\bigg|
%&= \bigg\vert \frac{(\sum_{i\in A} x_i)(\sum_{i\in I} y_i) - (\sum_{i\in I} x_i)(\sum_{i\in A} y_i)}{(\sum_{i\in I} x_i)(\sum_{i\in I} y_i)}\bigg\vert\\
&= \frac{|\sum_{i'\in A} \sum_{i\in I} x_{i'} y_i - \sum_{i'\in A} \sum_{i\in I} x_i y_{i'}|}{\sum_{i',i\in I} x_{i'} y_i}\\
&= \frac{|\sum_{i'\in A} \sum_{i\in A^c} x_{i'} y_i - \sum_{i'\in A} \sum_{i\in A^c} x_i y_{i'}|}{\sum_{i',i\in I} x_{i'} y_i}\\
&\leq \frac{\sum_{i'\in A} \sum_{i\in A^c} |x_{i'} y_i - x_i y_{i'}|}{\sum_{i',i\in I} x_{i'} y_i}
\eel
where $A^c = I \setminus A$.
Second, if $a_i \geq 0$, $y_i > 0$ for all $i\in I$, $A\subseteq I$, and $\sum_{i\in I} a_i y_i > 0$, then
\bel\label{eq:ratio_of_sums_inequality}
\frac{\sum_{i\in A} a_i x_i}{\sum_{i\in I} a_i y_i} = \frac{\sum_{i\in A} a_i (x_i/y_i)y_i}{\sum_{i\in I} a_i y_i} \leq \frac{\sum_{i\in A} a_i (\max_{j\in A} x_j/y_j)y_i}{\sum_{i\in I} a_i y_i} \leq\max_{i\in A} x_i/y_i.
\eel

\vspace{1em}
\begin{proof}{\bf of \cref{thm:consistent}}~\\
	In \cref{thm:consistent0}, we proved posterior consistency of the number of clusters in the case of $\epsilon = 0$, so we only need to show that $|\Pr_{\epsilon(n)}(T=t \mid y_{1:n}) - \Pr_{\epsilon = 0}(T=t \mid y_{1:n})| \to 0$ as $n\to\infty$. 
	We abbreviate $y = y_{1:n}$ to reduce notational clutter.
	First, using \cref{eq:postt}, for any integer $1\leq t\leq n$,
	\begin{align}\label{eq:bound_difference}
		& |\Pr_{\epsilon}(T=t \mid {y}) - \Pr_{\epsilon = 0}(T=t \mid {y})| \notag \\
		& =
		\Biggl| \frac { \sum_{\mathcal{A} \in \mathcal{H}_t(n)} \Pr ({y}\mid \mathcal{A}) \Pr_{\epsilon,n} (\mathcal{A}) } { \sum_{\mathcal{A}\in \cup_{t=1}^n \mathcal{H}_t(n)} \Pr ({y}\mid \mathcal{A})\Pr_{\epsilon,n} (\mathcal{A}) } 
		- \frac { \sum_{\mathcal{A} \in \mathcal{H}_t(n)} \Pr ({y}\mid \mathcal{A}) \Pr_{0,n} (\mathcal{A}) } { \sum_{\mathcal{A}\in\cup_{t=1}^n \mathcal{H}_t(n)} \Pr ({y}\mid \mathcal{A})\Pr_{0,n} (\mathcal{A}) } \Biggr| \notag\\
		&\stackrel{\textup{(a)}}{\leq} \frac {
			\sum_{\mathcal{A'} \in \mathcal{H}_t(n)} \sum_{\mathcal{A} \in\cup_{l\neq t} \mathcal{H}_l(n)}
			\Pr ({y}\mid \mathcal{A'}) \Pr ({y}\mid \mathcal{A})
			| \Pr_{\epsilon,n} (\mathcal{A'}) \Pr_{0,n} (\mathcal{A}) 
			-  \Pr_{0,n} (\mathcal{A'}) \Pr_{\epsilon,n} (\mathcal{A}) |
		}
		{ 
			\sum_{\mathcal{A'},\mathcal{A} \in \cup_{t=1}^n \mathcal{H}_t(n)} \Pr ({y}\mid \mathcal{A'}) \Pr ({y}\mid \mathcal{A}) \Pr_{\epsilon,n} (\mathcal{A'}) \Pr_{0,n} (\mathcal{A}) }\notag\\
		&\stackrel{\textup{(b)}}{\leq} \max_{\mathcal{A'} \in \mathcal{H}_t(n)} \max_{\mathcal{A} \in\cup_{l\neq t} \mathcal{H}_l(n)} \biggl|1-\frac{\Pr_{0,n} (\mathcal{A'}) \Pr_{\epsilon,n} (\mathcal{A})}{\Pr_{\epsilon,n} (\mathcal{A'}) \Pr_{0,n} (\mathcal{A})}\biggr|, 
	\end{align}
	where (a) and (b) are by \cref{eq:difference_of_ratios_inequality,eq:ratio_of_sums_inequality}, respectively. To ensure that the denominators in the preceding display are nonzero, there needs to exist at least one partition $\mathcal{A} \in \cup_{t=1}^n \mathcal{H}_t(n)$ such that $\Pr ({y}\mid \mathcal{A}) > 0$, and this is indeed the case since (1) with probability 1, $p(y\mid\phi_0) > 0$, and (2) we assume the support of the prior $\mathcal{G}(\theta)$ contains a neighborhood of each $\theta^0_k$ and the component density $f_\theta$ is continuous (with respect to $\theta$) at each $\theta^0_k$. 
	
	We use the same notation as in the proof of \cref{thm:prior}, where we have already proved that
	\begin{align}\label{eq:lowerbound}
		\frac{\Pr_{0,n}(\mathcal{A})}{\Pr_{\epsilon,n}(\mathcal{A})} 
		\leq  1 + \frac{\alpha n\epsilon}{\alpha+1-\alpha\epsilon n}
	\end{align}
	for any partition $\mathcal{A}$. Next, we construct an upper bound on the reciprocal, $\Pr_{\epsilon,n}(\mathcal{A})/\Pr_{0,n}(\mathcal{A})$, by upper bounding $V_j({\sigma}) / U_j({\sigma})$ for each $j$. We split the analysis into two cases: $j\leq t-1$ and $j=t$.
	
	Case 1: $j\leq t-1$.  This implies $g_{j+1}(\sigma)\geq t-j \geq 1$ since every cluster has at least one element.
	Letting $r$ denote the integer such that $0<\alpha-r\leq 1$ (or equivalently, $\max(\alpha-1,0)\le r< \alpha$),
	\begin{align*}
		& \frac{1}{\epsilon^\alpha} I_{\epsilon}( g_{j+1}(\sigma) + \alpha, n_{\sigma_j}+ 1 )\\
		& \leq \frac{1}{\epsilon^\alpha \mathrm{B}(g_{j+1}(\sigma) + \alpha, n_{\sigma_j}+ 1)}\int_{0}^\epsilon x^{g_{j+1}(\sigma) + \alpha - 1}\,d x \\
		& = \frac{\Gamma(g_j(\sigma)+\alpha+1) \epsilon^{g_{j+1}(\sigma)}} {\Gamma(g_{j+1}(\sigma) + \alpha)\Gamma(n_{\sigma_j}+1)(g_{j+1}(\sigma)+\alpha) }\\
		& =   \epsilon^{g_{j+1}(\sigma)}\frac{(g_j(\sigma)+\alpha)(g_j(\sigma)+\alpha-1)\cdots(\alpha-r)}{(g_{j+1}(\sigma)+\alpha)(g_{j+1}(\sigma)+\alpha-1)\cdots(\alpha-r)}
		\frac{1}{n_{\sigma_j}!}\\
		& = \epsilon^{g_{j+1}(\sigma)}\Bigg(\prod_{m=0}^{g_{j+1}(\sigma)+r} \frac{n_{\sigma_j}+\alpha-r+m}{\alpha-r+m}\Bigg)
		\Bigg(\prod_{m=1}^{n_{\sigma_j}} \frac{m+\alpha-r-1}{m}\Bigg) \\
		%& =  \frac{g_j(\sigma)+\alpha}{g_{j+1}(\sigma)+\alpha}\frac{g_{\sigma_{j}}+\alpha-1}{g_{j+1}(\sigma)+\alpha-1}\cdots\frac{n_{\sigma_j}+\alpha-r}{\alpha-r} \cdot \frac{n_{\sigma_j}+\alpha-r-1}{n_{\sigma_j}}\frac{n_{\sigma_j}+\alpha-r-2}{n_{\sigma_j}-1}\cdots\frac{\alpha-r}{1} \epsilon^{g_{j+1}(\sigma)} \\
		& \stackrel{\textup{(a)}}{\leq}  \epsilon^{g_{j+1}(\sigma)} \biggl(1 + \frac{n_{\sigma_j}}{\alpha-r}\biggr)^{g_{j+1}(\sigma)+r+1} \\
		& \stackrel{\textup{(b)}}{\leq}  \epsilon^{t-j} \biggl(1 + \frac{n}{\alpha-r}\biggr)^{t-j+r+1} 
	\end{align*}
	for all $n$ sufficiently large, where (a) results from $(n_{\sigma_j}+\alpha-r+m)/(\alpha-r+m)\leq (n_{\sigma_j}+\alpha-r)/(\alpha-r)$ for $m=0,1,\dots,g_{j+1}(\sigma)+r$
	and $(m+\alpha-r-1)/m\leq 1$ for $m=1,\dots,n_{\sigma_j}$, and (b) uses $n_{\sigma_j}\leq n$, 
	$\epsilon(1+n/(\alpha-r))<1$ for all $n$ sufficiently large since $\epsilon =\epsilon(n) = o(1/n)$, and $g_{j+1}(\sigma)\geq t-j$.

	Combining this with
	\begin{equation}\label{eq:upb1}
		\frac{g_j(\sigma) + \alpha (1 - \tilde{p})} {g_j(\sigma) + \alpha (1 - \tilde{p}) (1 - \epsilon^{g_j(\sigma)}) } = 1 + \frac{ \alpha (1 - \tilde{p})\epsilon^{g_j(\sigma)} } {g_j(\sigma) + \alpha (1 - \tilde{p}) (1 - \epsilon^{g_j(\sigma)})} \leq 1+\alpha(1-\tilde{p})\epsilon ,
	\end{equation}
	we have
	\[
	\frac{V_j({\sigma})}{U_j({\sigma})} 
	\leq \bigg(1 + \frac{1-\tilde{p}}{\tilde{p}}  \epsilon^{t-j} \Big(1 + \frac{n}{\alpha-r}\Big)^{t-j+r+1} \bigg) \big(1 + \alpha (1 - \tilde{p})\epsilon\big).
	\]
	
	Case 2: $j=t$. We have
	\[
	\frac{V_t({\sigma})}{U_t({\sigma})} 
	\leq 1 + \alpha (1 - \tilde{p})\epsilon
	\]
	since \cref{eq:upb1} holds when $j=t$ and
	\[
	\frac{I_\epsilon(\alpha,n_{\sigma_t}+1)}{\epsilon^\alpha} 
	\leq \frac{1}{\epsilon^\alpha\mathrm{B}(\alpha,n_{\sigma_t}+1)}\int_0^\epsilon x^{\alpha-1}\,d x 
	= \frac{1}{\alpha\mathrm{B}(\alpha,n_{\sigma_t}+1)}.
	\] 
	
	Thus, using the same expression as in \cref{eq:pmfratio},
	\begin{align}\label{eq:upperbound}
		&\frac{\Pr_{\epsilon,n}(\mathcal{A})}{\Pr_{0,n}(\mathcal{A})}
		\leq  \max_{ {\sigma}\in S_t} \prod_{j=1}^t \frac{V_j({\sigma})} {U_j({\sigma})}\notag\\
		&\leq \big(1 + \alpha (1 - \tilde{p})\epsilon\big)^t \prod_{j=1}^{t-1} \bigg(1 + \frac{1-\tilde{p}}{\tilde{p}}  \epsilon^{t-j} \Big(1 + \frac{n}{\alpha-r}\Big)^{t-j+r+1} \bigg) \notag\\
		&\leq \big(1 + \alpha(1-\tilde{p}) \epsilon\big)^n \bigg(1 + \frac{1-\tilde{p}}{\tilde{p}}  \epsilon^{2} \Big(1 + \frac{n}{\alpha-r}\Big)^{3+r} \bigg)^n \bigg(1 + \frac{1-\tilde{p}}{\tilde{p}}  \epsilon \Big(1 + \frac{n}{\alpha-r}\Big)^{2+r} \bigg)
	\end{align}
	since $t=|\mathcal{A}|\leq n$ and $\epsilon(1+n/(\alpha-r)) < 1$ for $n$ sufficiently large, because $\epsilon=\epsilon(n)=o(1/n)$.
	
	Since $\epsilon(n)=o(1/n^{2+r})$, the upper bound on $\Pr_{\epsilon(n),n}(\mathcal{A})/\Pr_{0,n}(\mathcal{A})$ in \cref{eq:upperbound} converges to $1$ as $n\to\infty$. Meanwhile, \cref{eq:lowerbound} provides a lower bound that also converges to $1$.
	Since these upper and lower bounds depend only on $\alpha$, $r$, $\tilde{p}$, $\epsilon$, and $n$, they hold uniformly over all $\mathcal{A}\in \cup_{t=1}^n \mathcal{H}_t(n)$.  Hence, $\Pr_{\epsilon(n),n}(\mathcal{A})/\Pr_{0,n}(\mathcal{A})\to 1$ uniformly over $\mathcal{A}$, as $n\to\infty$.
	Therefore, along with \cref{eq:bound_difference}, this implies that
	\[
	|\Pr_{\epsilon(n)}(T=t \mid {y}) - \Pr_{\epsilon = 0}(T=t \mid {y})|\leq
	\max_{\mathcal{A'} \in \mathcal{H}_t(n)} \max_{\mathcal{A} \in\cup_{l\neq t} \mathcal{H}_l(n)} \biggl|1-\frac{\Pr_{0,n} (\mathcal{A'}) \Pr_{\epsilon(n),n} (\mathcal{A})}{\Pr_{\epsilon(n),n} (\mathcal{A'}) \Pr_{0,n} (\mathcal{A})}\biggr|
	\longrightarrow 0
	\]
	as $n\to\infty$.
\end{proof}

\vspace{1em}
\begin{proof}{\bf of Lemma \ref{lemma:dp0_inconsistency}}~\\
	To study the posterior distribution of the number of clusters $p(T = t \mid y_{1:n})$, we will focus on
	\bel\label{eq:marginal_posterior}
	p(y_{1:n},T=t) & = \sum_{\mathcal{A}\in \mathcal{H}_t(n)}  p(y_{1:n} \mid \mathcal{A}) p(\mathcal{A})
	\eel
	where $\mathcal{A} = \{A_1,\dots,A_t\}$, $p ({y_{1:n}} \mid \mathcal{A}) = \prod_{A\in\mathcal{A}} m(y_A)$, $y_A = (y_i:i\in A)$, and $m(y_A) = \int_{\Theta}\bigl( \prod_{i\in A} f_{\theta}(y_i) \bigr)\, d\mathcal{G}(\theta)$, as described at \cref{eq:postt}.
	In this lemma, $f_\theta(y_i) = \No(y_i\mid \theta,1)$ and
	\bel\label{eq:marginal_likelihood}
	m(y_A) = \frac{1}{\sqrt{|A|+1}} f_0(y_A) \exp\bigg(\frac{(\sum_{i\in A} y_i)^2}{2(|A|+1)}\bigg),
	\eel
	where $f_0(y_A) = \prod_{j\in A} \No(y_j\mid 0,1)$.
	Under the Dirichlet process prior, the probability mass function on partitions is \(p(\mathcal{A}) = \frac{\alpha^t}{\alpha^{(n)}}\prod_{i=1}^t(|A_i|-1)!\) where \(\alpha^{(n)} = \alpha(\alpha+1)\cdots(\alpha+n-1)\). Hence,
	\be
	\frac{p(y_{1:n}, T = 2)}{p(y_{1:n}, T = 1)} 
	&\stackrel{\textup{(a)}}{=} \frac{\sum_{\mathcal{A} = \{A_1,A_2\} \in \mathcal{H}_2(n)} p(\mathcal{A})m(y_{A_1})m(y_{A_2})}
	{\mathbb{P}(\mathcal{A}=\big\{\{1,\dots,n\}\big\})m(y_{1:n})}\\
	&\stackrel{\textup{(b)}}{=} \sum_{\mathcal{A}= \{A_1,A_2\} \in \mathcal{H}_2(n)} 
	\biggl[
	\frac {{\alpha^2(|A_1|-1)!(|A_2|-1)!}} {{\alpha (n-1)!}} \times \frac{\sqrt{n+1}}{\sqrt{|A_1|+1} \sqrt{|A_2|+1}}\\
	& ~~~~~~~ \times \exp\bigg( \frac{(\sum_{i\in A_1} y_i)^2}{2(|A_1|+1)} \bigg) \exp\bigg( \frac{(\sum_{i\in A_2} y_i)^2}{2(|A_2|+1)} \bigg)  \exp\bigg(-\frac{(\sum_{i=1}^n y_i)^2}{2(n+1)} \bigg)
	\biggr]\\
	%\stackrel{\textup{(c)}}{\leq}{} & \frac{\alpha\sqrt{n+1} }{(n-1)!} \sum_{\mathcal{A} =\{A_1,A_2\} \in \mathcal{H}_2(n)} \frac{(|A_1|-1)!(|A_2|-1)!}{\sqrt{|A_1|+1}\sqrt{|A_2|+1}}
	%\exp\bigg(\frac{|A_1|\sum_{i\in A_1}y_i^2}{2(|A_1|+1)}+\frac{|A_2|\sum_{i\in A_2}y_i^2}{2(|A_2|+1)}\bigg)\\
	&\stackrel{\textup{(c)}}{\leq} \frac{\alpha\sqrt{n+1} }{(n-1)!} \sum_{\mathcal{A} \in \mathcal{H}_2(n)}     \frac{(|A_1|-1)!(|A_2|-1)!}{\sqrt{|A_1|+1}\sqrt{|A_2|+1}} \exp\bigg(\frac{\sum_{i\in A_1}y_i^2}{2}+\frac{\sum_{i\in
			A_2}y_i^2}{2}\bigg)\\
	&\stackrel{\textup{(d)}}{\leq} \frac{\alpha\sqrt{n+1} }{2(n-1)!} \exp\bigg(\frac{\sum_{i=1}^n y_i^2}{2} \bigg) \sum_{\mathcal{A} \in \mathcal{H}_2(n)} (|A_1|-1)!(|A_2|-1)!\\
	&= \frac{\alpha\sqrt{n+1} }{2(n-1)!} \exp\bigg(\frac{\sum_{i=1}^n y_i^2}{2} \bigg) \sum_{i=1}^{n-1} \frac{1}{2} {n\choose i} (i-1)!(n-i-1)!\\
	&= \frac{\alpha\sqrt{n+1} }{2} \exp\bigg(\frac{\sum_{i=1}^n y_i^2}{2} \bigg) \sum_{i=1}^{n-1} \frac{n}{2i(n-i)}\\
	&= \frac{\alpha\sqrt{n+1} }{2} \exp\bigg(\frac{\sum_{i=1}^n y_i^2}{2} \bigg) \biggl(1+\frac{1}{2}+\cdots+\frac{1}{n-1}\biggr)\\
	&\leq \frac{\alpha\sqrt{n+1} }{2} \exp\bigg(\frac{\sum_{i=1}^n y_i^2}{2} \bigg) \big(\log(n-1)+1\big)\\
	\ee
	where (a) is using \cref{eq:marginal_posterior} for both numerator and denominator, (b) is using \cref{eq:marginal_likelihood} and the probability mass function of $\mathcal{A}$, (c) follows from $(\sum_{j=1}^n y_j)^2 \geq 0$ and Jensen's inequality, (d) is using $A_1\cup A_2 = \{1,\dots,n\}$ and both $|A_1|$ and $|A_2|$ are greater than or equal to $1$, and the last inequality is induced from $1/k \leq \log(k) - \log(k-1)$ for $k=2,\dots,n-1$.

	By the strong law of large numbers, $$\frac{1}{n}\sum_{i=1}^n y_i^2 \xrightarrow[n\to\infty]{\mathrm{a.s.}} E(y_1^2) = 1+\kappa^2/2.$$
	Let $\delta = C - (1/2 + \kappa^2/4)$, where by the assumption of the theorem, $C > 1/2 + \kappa^2/4$ such that $\alpha = \alpha(n) = o(\exp(-C\, n))$. Then, almost surely, for all $n$ sufficiently large, $\frac{1}{2 n}\sum_{i=1}^n y_i^2 \leq 1/2 + \kappa^2/4 + \delta/2 = C - \delta/2$. Hence, almost surely,
	\be
	\limsup_{n\to\infty} \frac{p(y_{1:n}, T = 2)}{p(y_{1:n}, T = 1)} 
	& \leq \frac{\alpha(n)\sqrt{n+1} }{2} \exp\big(n (C - \delta/2)\big) \big(\log(n-1)+1\big) \\
	& = \big(\alpha(n) \exp(C\,n)\big) \exp(-n\delta/2) \frac{\sqrt{n+1} }{2} \big(\log(n-1)+1\big) 
	\longrightarrow 0
	\ee
	as $n\to\infty$. Therefore, we have the conclusion,
	\[
	\begin{aligned}
		p(T=2 \mid y_{1:n})  = \frac{p(y_{1:n},T=2)}{\sum_{t=1}^{\infty} p(y_{1:n},T=t)} \leq \frac{p(y_{1:n},T=2)}{p(y_{1:n},T=1)} \stackrel{\text{a.s.}}{\longrightarrow} 0.
	\end{aligned}
	\]
	
\end{proof}

\section{Additional Simulation Results}
\label{sec:additional_sim}

{In this section, we first provide the simulation results when the component distribution is the bivariate Gaussian distribution. Then we show a comparison between the quasi-Bernoulli mixture and the Dirichlet process mixture with $\alpha(n)\to 0$. Finally, we provide the information of convergence diagnostics and running time of the algorithm.}

\subsection{Simulation Results with Bivariate Gaussian Mixtures}\label{subsec:sim_multiGau}

\begin{figure}
	\centering
	\includegraphics[width=1\linewidth]{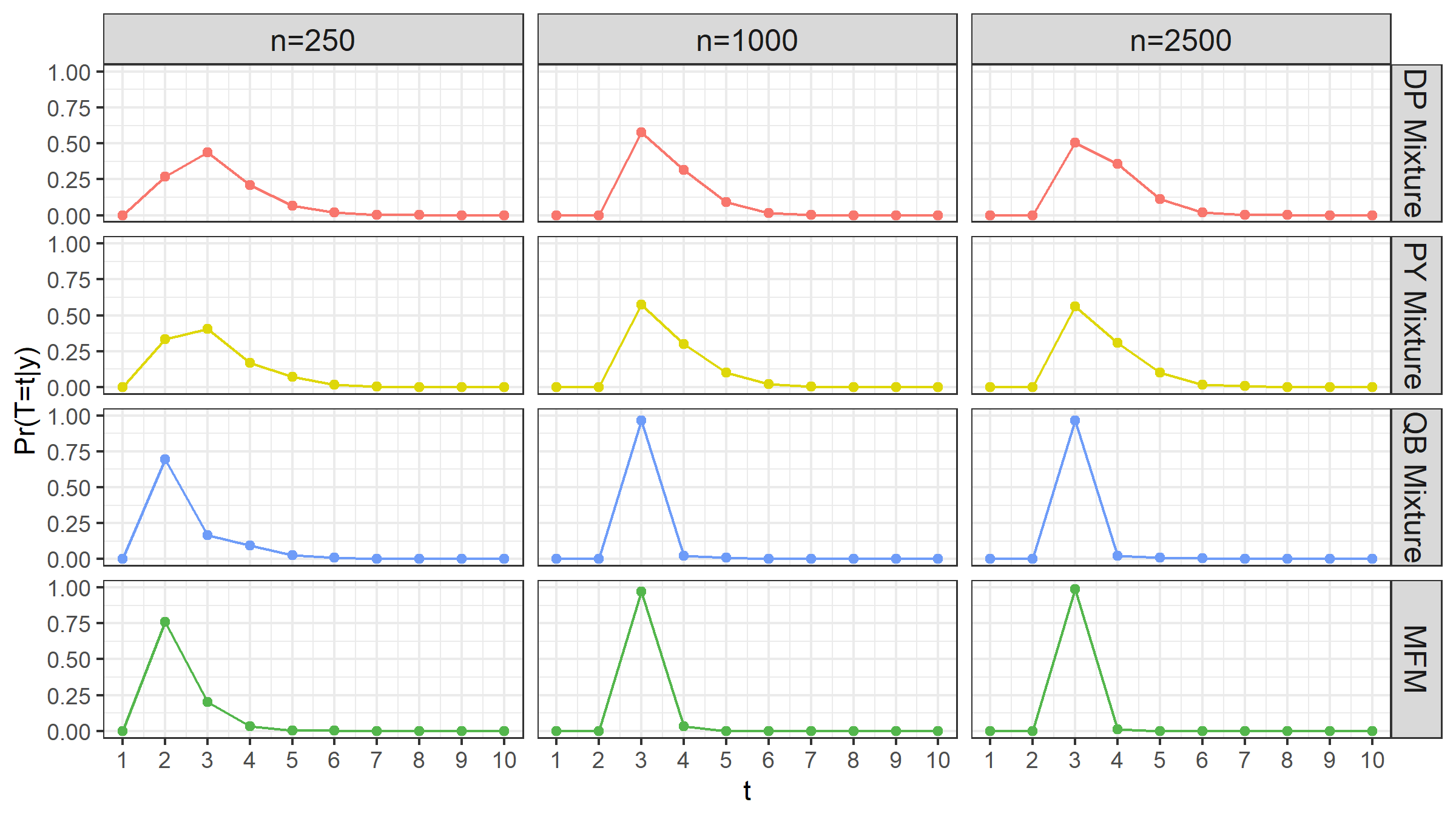}
	\caption{Posterior distribution on number of clusters for data generated from a three-component Gaussian mixture in $\mathbb{R}^2$. Using the quasi-Bernoulli mixture model, the posterior probabilities of $T$ converge to a point mass at $k_0=3$ for large $n$. However, the posterior distributions of the calibrated Dirichlet process mixture model and Pitman--Yor process mixture model do not converge to a point mass at $k_0=3$.}\label{fig:2D}
\end{figure}

\Cref{fig:2D} plots the posterior distribution of the number of clusters $T$ at each $n$. Under the quasi-Bernoulli mixture model (shown in blue), the posterior of $T$ converges to a point mass at the true number of components ($k_0 = 3$) as $n$ grows, in accordance with our theory.  On the other hand, the posterior distributions of $T$ with Dirichlet process mixture model and the Pitman--Yor process mixture model fail to converge to a point mass at the true number of components.

\begin{figure}
	\begin{subfigure}[t]{0.49\textwidth}
		\centering
		\includegraphics[width=1\linewidth]{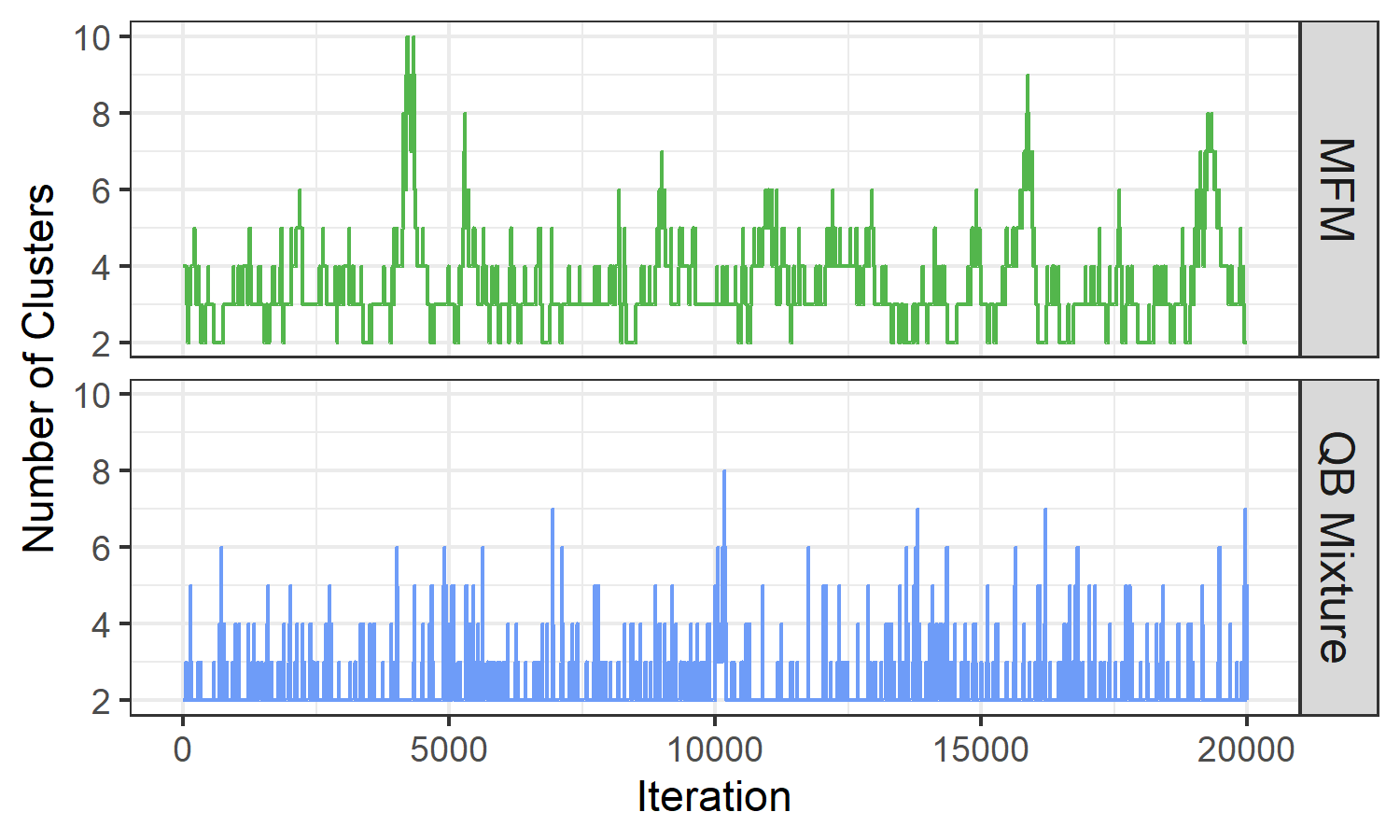}
		\caption{Trace on number of clusters}
	\end{subfigure}
	\begin{subfigure}[t]{0.49\textwidth}
		\centering
		\includegraphics[width=1\linewidth]{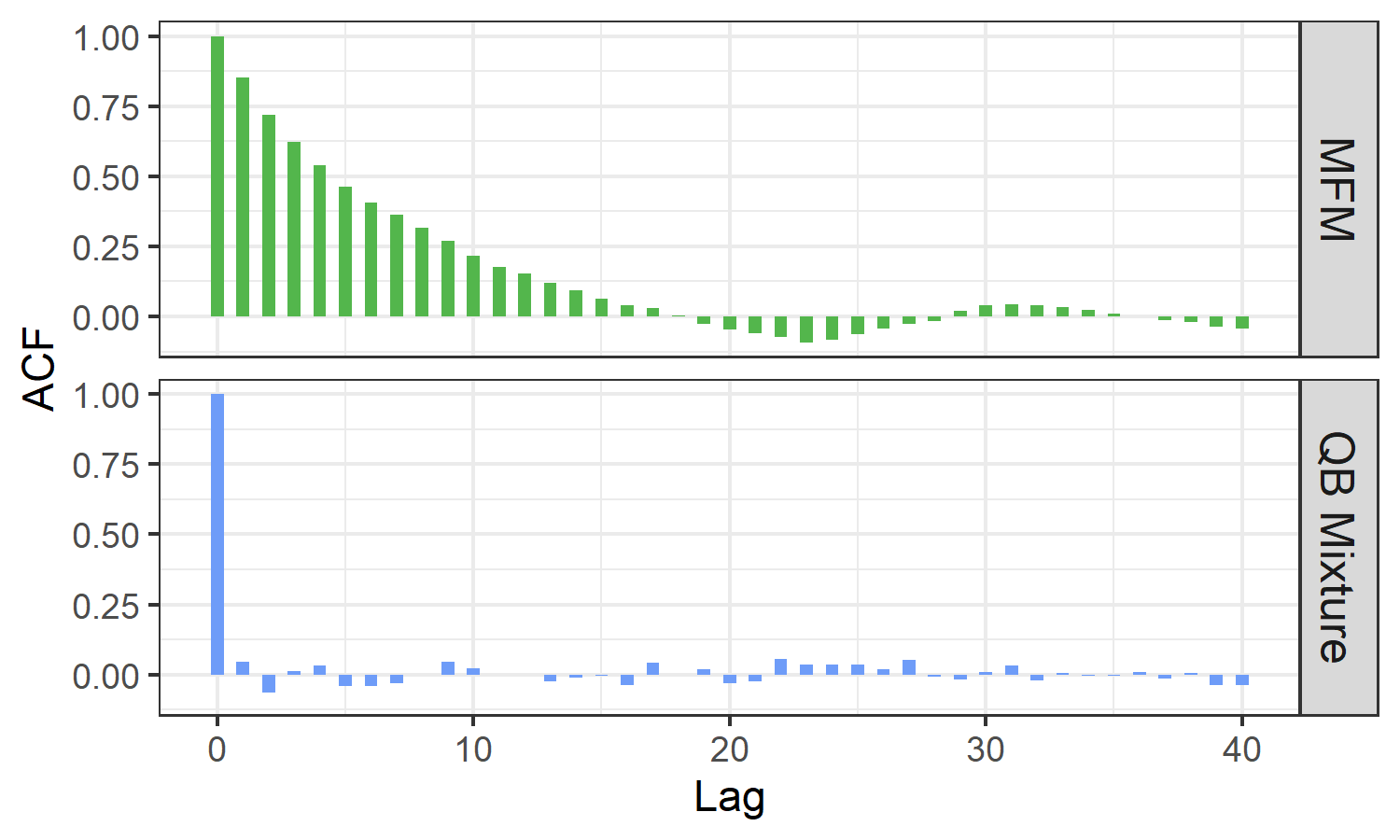}
		\caption{Auto-correlation (based on thinning at $25$)} 
	\end{subfigure}
	\centering
	\caption{The trace of the Markov chain on \(T\) and auto-correlation functions for bivariate Gaussian mixture data with sample size \(250\). Quasi-Bernoulli mixture model shows much better mixing in the Markov chain, compared to the MFM model. We discard the first 5,000 iterations as burn-ins and record the following 20,000 samples.}
	\label{fig:computing_perf2}
\end{figure}

{As shown in Figure~\ref{fig:computing_perf2},
	the MFM model suffers from slow mixing with high auto-correlation even after thinning (effective sample size $15.3\%$ on average of five experiments with sample size $250$); whereas the quasi-Bernoulli mixture model quickly shows a much faster drop in the auto-correlation within a few lags  (effective sample size $57.2\%$).}

\subsection{Simulations on the Dirichlet Process Mixture with $\alpha(n)\to 0$}
\label{subsec:dp_sim}
\begin{figure}
	\centering
	\includegraphics[width=1\linewidth]{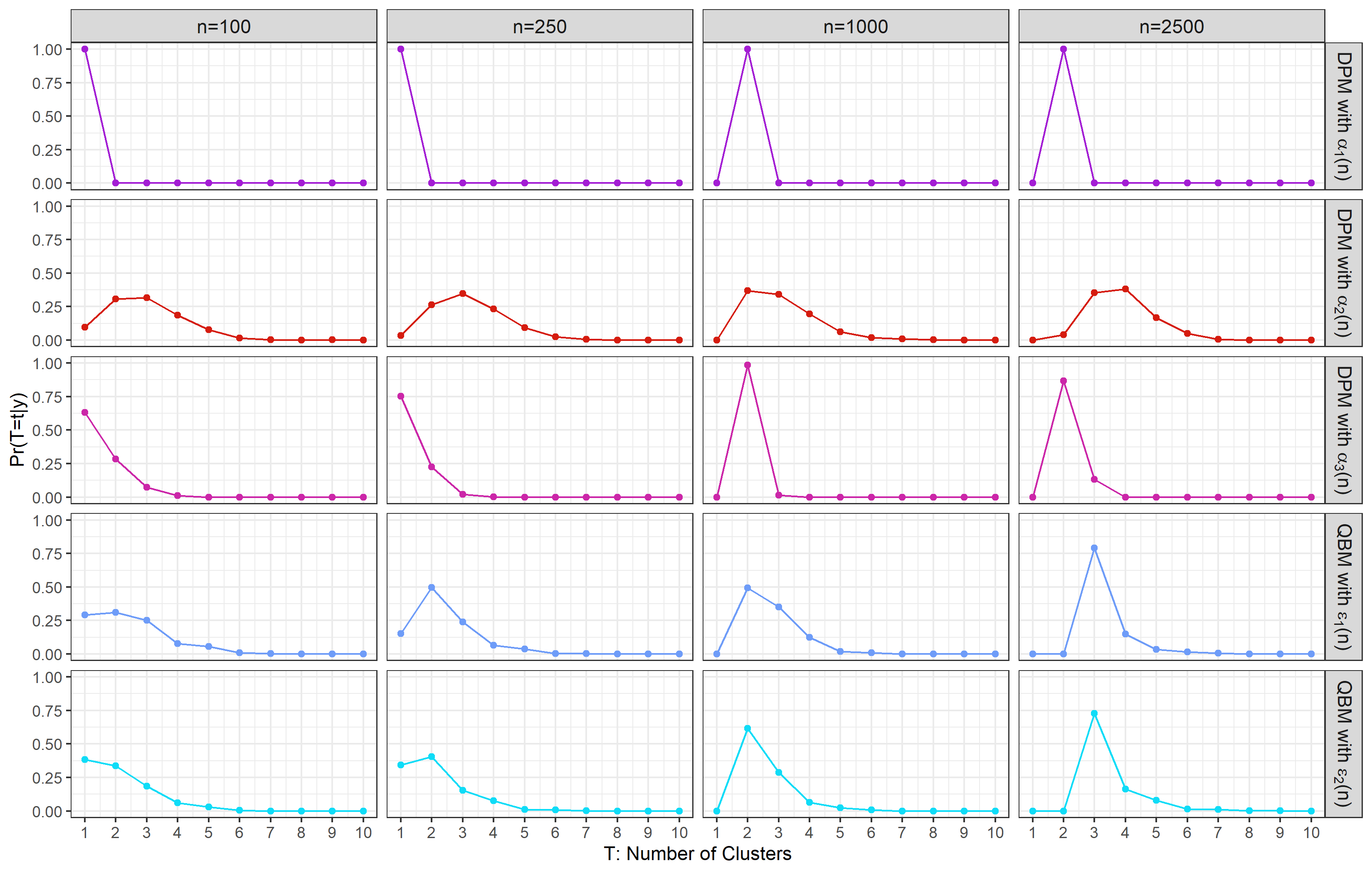}
	\caption{Posterior distribution on number of clusters for data generated from a three-component univariate Gaussian mixture. 
	}\label{fig:1Dano}
\end{figure}

{To compare the quasi-Bernoulli mixture model with the Dirichlet process mixture model under different rates of $\alpha(n)\to 0$, we conduct additional experiments with univariate Gaussian mixtures. The experimental settings are similar to the ones in \cref{subsec:uniGau}, except that we  generate data from mixtures with smaller distances between the component centers: $0.3\No(-2, 1^2) + 0.4\No(0, 1^2) + 0.3\No(2, 1^2)$ under sample sizes $n\in\{100,250,1000,2500\}$. The purpose is to examine if each model shows the trend of converging to the ground-truth $T=k_0$ as $n$ increases, when the component distribution $\mathcal{F}$ has an unbounded support and clusters have large overlaps.
	
	For Dirichlet process mixture models, we use three rates $\alpha_1(n) = \exp(-n/10)$, $\alpha_2(n) = 4/\log(n)$ and $\alpha_3(n)=20/n$. For quasi-Bernoulli mixture models, we use two rates   $\epsilon_1(n) = n^{-2.1}$ and $\epsilon_2(n) = n^{-3.1}$. \cref{fig:1Dano} shows the posterior distributions of the number of clusters. 
	
	This empirical result suggests that one may be able to obtain posterior consistency on estimating $T$ for the Dirichlet process mixture model with general $\mathcal{F}$, by choosing an $\alpha\to 0$ faster than $4/\log(n)$ but slower than $20/n$, although the theory remains an open question. On the other hand, the quasi-Bernoulli mixture models show almost no difference in the trend of convergence. This is as expected, since  both $n^{-2.1}$ and $n^{-3.1}$ satisfy the rate condition that guarantees the consistency on estimating $T$.
}

\subsection{Convergence Diagnostics and Timing Information}

{We use the Markov chain sample of $T$ (the number of clusters) for convergence diagnosis. We choose $T$ because it is often the slowest-changing variable.
	As shown in \cref{fig:computing_perf}, the auto-correlations for $T$ show a quick drop to insignificant level, within as few lags after thinning at $50$. For each experiment, we run multiple chains from 5 randomly initialized points, and compute the $\hat{R}$ statistic \citep{gelman1992inference}. All of the experiments get $\hat{R}$ close to $1$, which means that the Markov chains have converged. 
	% For example, for the univariate Gaussian data with sample size 250, the $\hat{R}=1.07$ and its confidential set has upper bound $1.18$; when the sample size increases to 1,000, $\hat{R}=1.11$ with upper bound $1.27$. For the data application, we get $\hat{R}=1.03$ and an upper bound $1.12$.
}

{We provide the timing information of our posterior sampling algorithms. The algorithms are implemented in R, and run on a 4.0 GHz processor. For the model with univariate Gaussian components, each iteration costs $0.0047,0.0060,0.0101,0.0350,0.0775$ seconds for the sample size $50,100,250,1000,2500$, respectively. For the bivariate Gaussian case, the algorithm runs $0.0029,0.0236,0.0945$ seconds for each iteration for sample size $250,1000,2500$, respectively. When the component distribution is the Laplace distribution, the algorithm takes $0.0055,0.1018$, $0.2269,0.3083$ seconds for each iteration for sample size $50, 200,500,1000$, respectively. For the network model used in the data application, each iteration takes around $0.3$ seconds.}

\section{Other Useful Results}

\begin{figure}[!h]
	\centering
	\begin{subfigure}[t]{0.32\textwidth}
		\centering
		\includegraphics[width=1\linewidth]{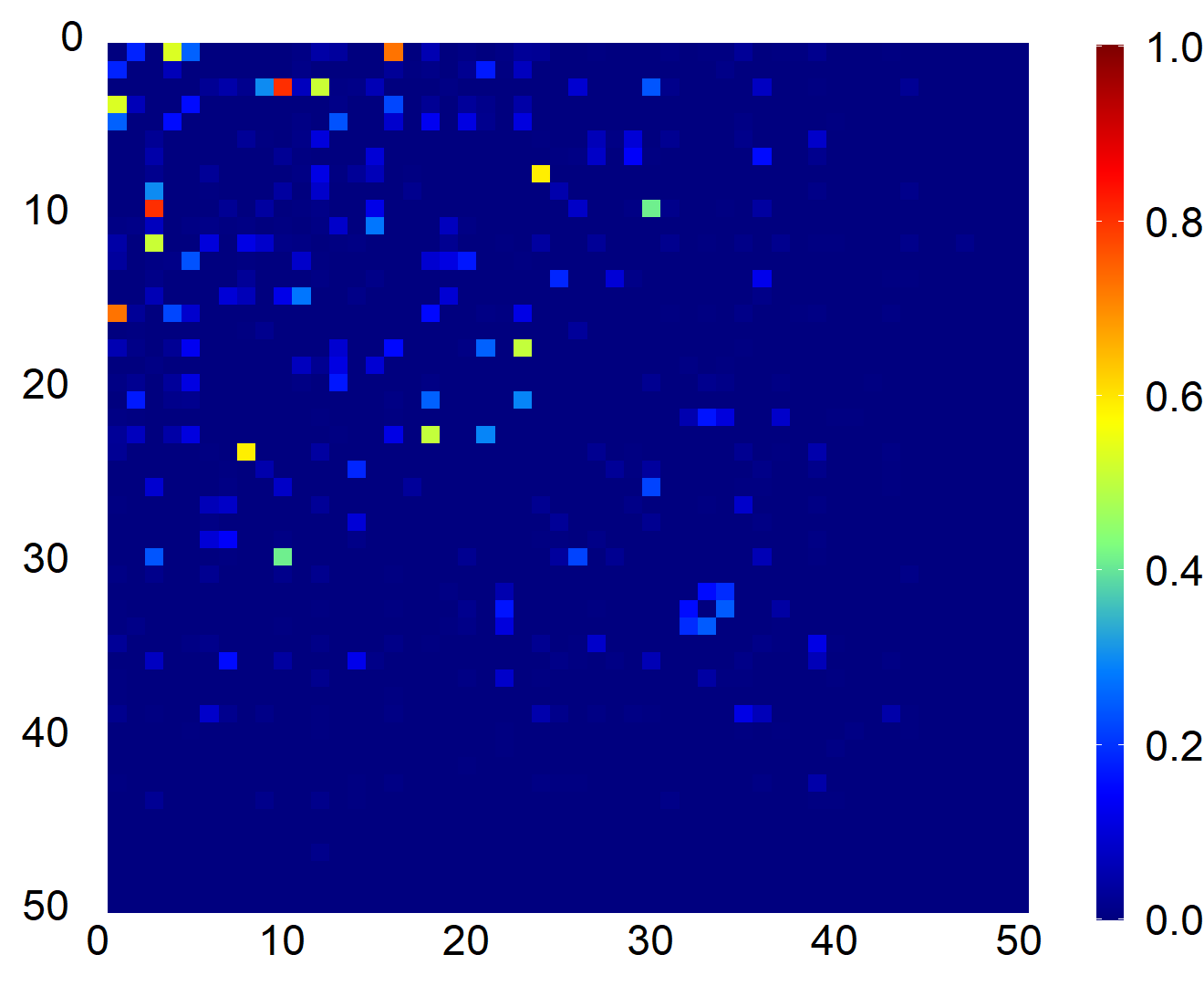}
		\caption{Group 1 ($36.6\%$ of the subjects)}
	\end{subfigure}
	\begin{subfigure}[t]{0.33\textwidth}
		\centering
		\includegraphics[width=0.97\linewidth]{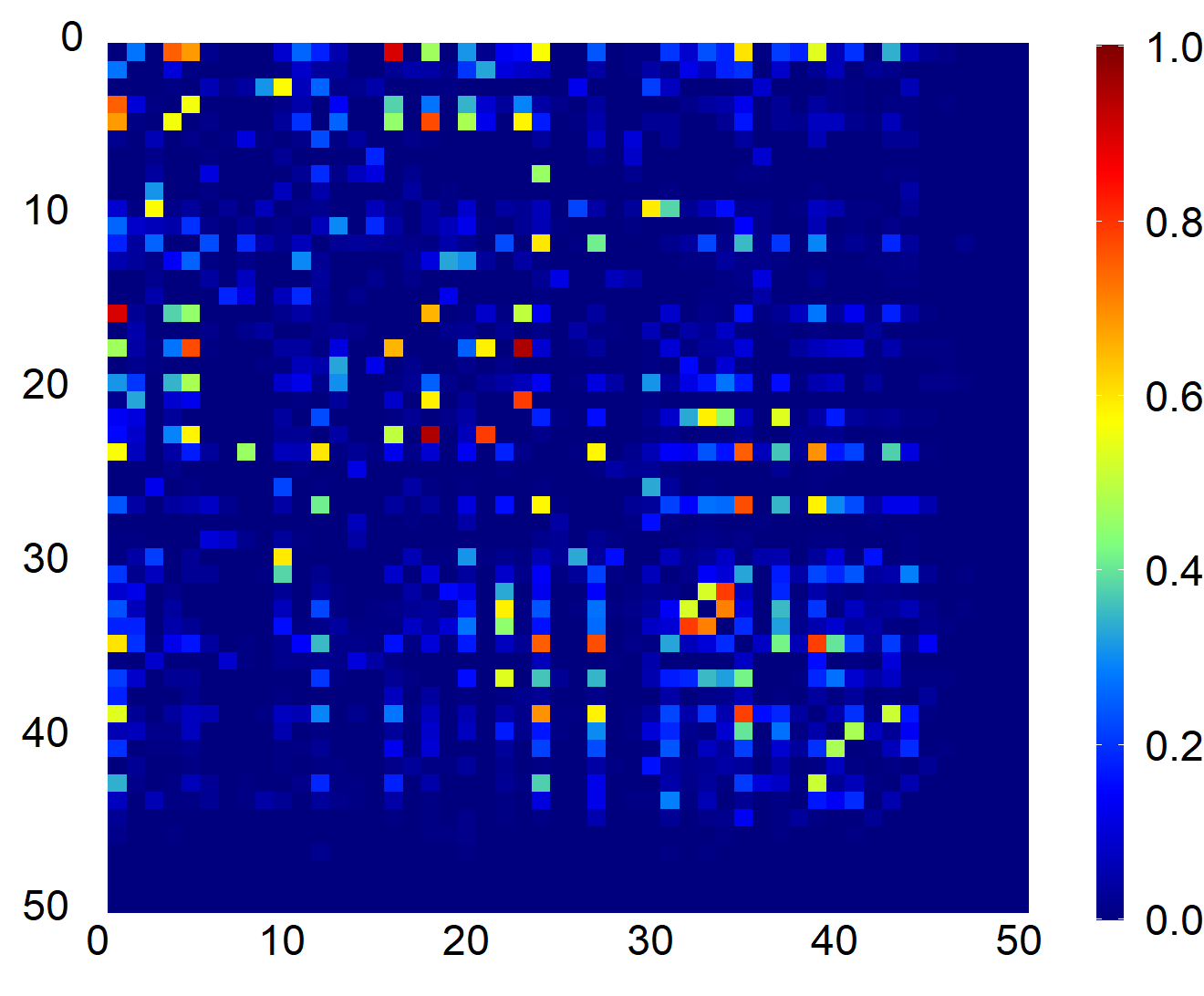}
		\caption{Group 2 ($20.6\%$ of the subjects)}
	\end{subfigure}
	\begin{subfigure}[t]{0.32\textwidth}
		\centering
		\includegraphics[width=1\linewidth]{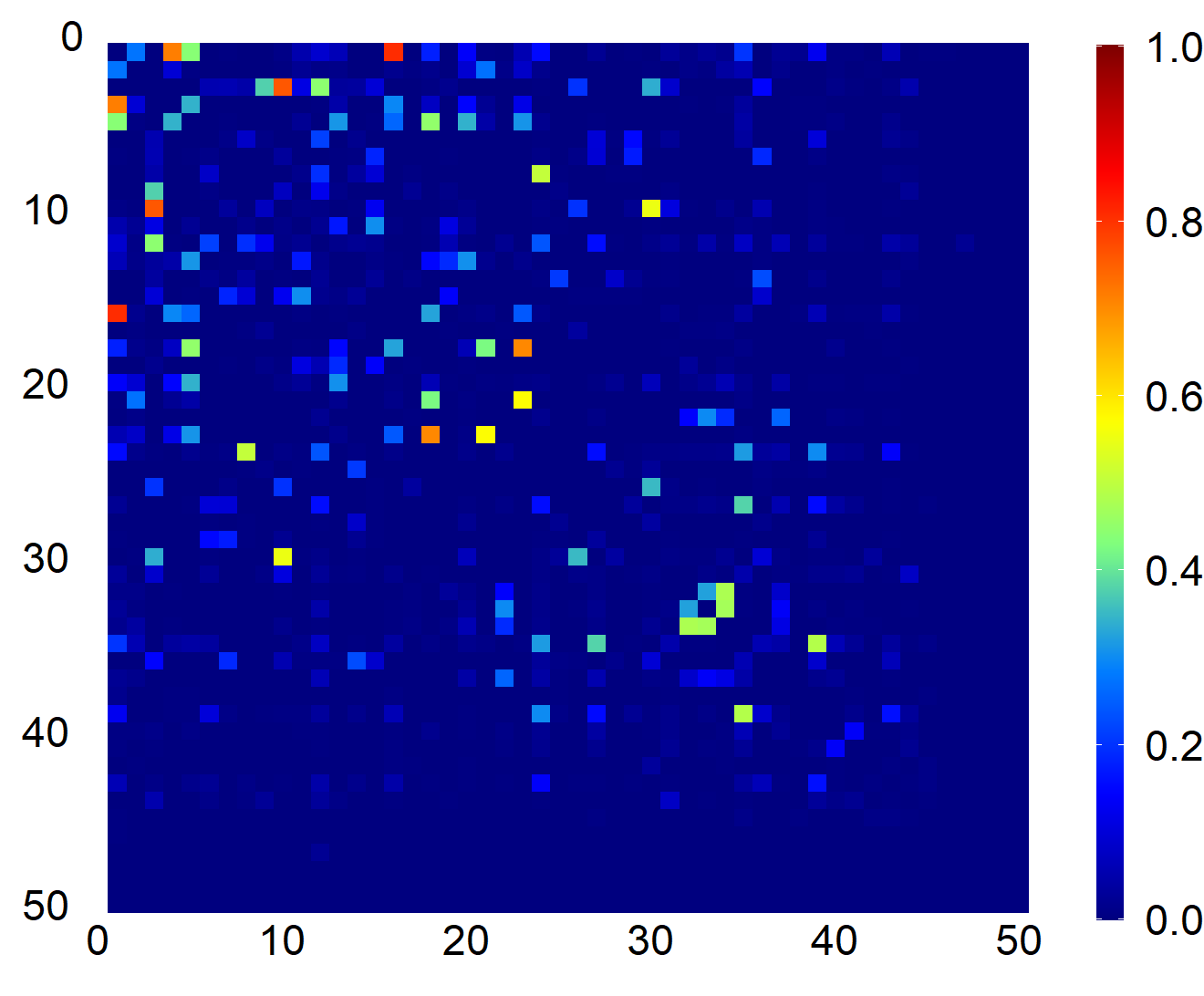} 
		\caption{Group 3 ($42.8\%$ of the subjects)}          
	\end{subfigure}
	\caption{The MAP estimation of the mean of each Gaussian component under the mixture of factor analyzers model. 
		\label{fig:fmri-mfa}}
\end{figure}

\begin{table}[!h]
	\centering
	\begin{tabular}{ccrccccrccr}
		\toprule
		& \multicolumn{2}{c}{Dirichlet process} &  & \multicolumn{4}{c}{Pitman--Yor process} & & \multicolumn{2}{c}{quasi-Bernoulli process}\\
		\cmidrule{2-3} \cmidrule{5-8} \cmidrule{10-11}
		& $\alpha$ & $\mathbb{E}(T)$ &  & $\alpha$ & $d$ & $\mathbb{E}(T)$ & $\text{Var}(T)$ &  & $\mathbb{E}(T)$ & $\text{Var}(T)$\\
		\midrule
		$50$  & $0.71$ & $3.62$ & & $0.48$ & $0.09$ & $3.62$ & $3.07$ & & $3.64$ & $3.09$\\
		$100$ & $0.69$ & $4.04$ & & $0.46$ & $0.09$ & $4.10$ & $3.95$ & & $4.06$ & $3.97$\\
		$250$ & $0.67$ & $4.58$ & & $0.38$ & $0.10$ & $4.56$ & $5.39$ & & $4.59$ & $5.40$\\
		$1000$& $0.63$ & $5.25$ & & $0.29$ & $0.11$ & $5.25$ & $8.25$ & & $5.28$ & $7.92$\\
		$2500$& $0.61$ & $5.69$ & & $0.29$ & $0.10$ & $5.74$ & $9.51$ & & $5.69$ & $9.79$\\
		\bottomrule
	\end{tabular}
	\caption{Settings of the hyper-parameters for the Dirichlet processes and the Pitman--Yor processes when sample sizes $n\in\{50,200,500,1000,2500\}$. We match the expectations of number of clusters $T$ under the three priors, and also make the variances of $T$ close under the Pitman-Yor process prior and the quasi-Bernoulli process prior. Each expectation and variance is approximated based on $2\times 10^5$ samples from the prior.}
	\label{tab:hyper-parameter}
\end{table}

\begin{figure}[!h]
	\centering
	\includegraphics[width=.7\linewidth]{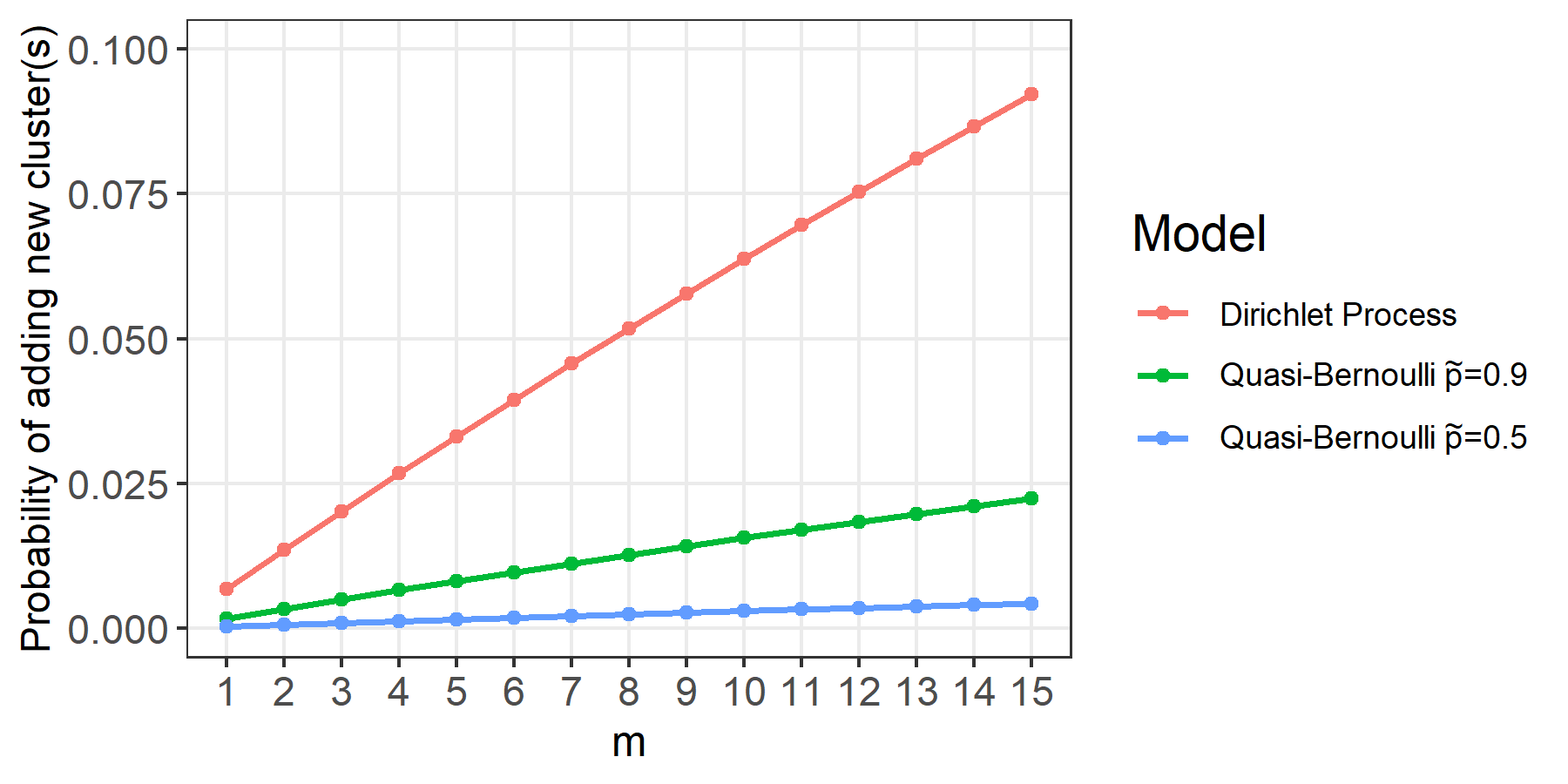}
	\caption{The probability of adding one or more new clusters for $m$ future data points ($n=100$). All of the parameters are chosen to be the same as \cref{fig:comp_w_dp} except for $\epsilon=\epsilon(n,m)=1/(n+m)^5$. The result does not change much.
		Under the prior, the Dirichlet process exhibits rapid growth in this probability, favoring the creation of additional clusters \textit{a priori}. Meanwhile, the quasi-Bernoulli process exhibits much slower growth of this probability.
		\label{fig:comp_w_dp2} }
\end{figure}

\newpage

\bibliographystyle{plainnat}

\bibliography{reference}

\begin{thebibliography}{30}
\providecommand{\natexlab}[1]{#1}
\providecommand{\url}[1]{\texttt{#1}}
\expandafter\ifx\csname urlstyle\endcsname\relax
  \providecommand{\doi}[1]{doi: #1}\else
  \providecommand{\doi}{doi: \begingroup \urlstyle{rm}\Url}\fi

\bibitem[Ascolani et~al.(2022)Ascolani, Lijoi, Rebaudo, and
  Zanella]{ascolani2022clustering}
Filippo Ascolani, Antonio Lijoi, Giovanni Rebaudo, and Giacomo Zanella.
\newblock {Clustering Consistency with Dirichlet Process Mixtures}.
\newblock \emph{arXiv preprint arXiv:2205.12924}, 2022.

\bibitem[Blackwell and MacQueen(1973)]{blackwell1973ferguson}
David Blackwell and James~B MacQueen.
\newblock {Ferguson Distributions via P{\'o}lya Urn Schemes}.
\newblock \emph{{The Annals of Statistics}}, 1\penalty0 (2):\penalty0 353--355,
  1973.

\bibitem[Cai et~al.(2021)Cai, Campbell, and Broderick]{cai2021finite}
Diana Cai, Trevor Campbell, and Tamara Broderick.
\newblock {Finite Mixture Models do not Reliably Learn the Number of
  Components}.
\newblock In \emph{International Conference on Machine Learning}, pages
  1158--1169. PMLR, 2021.

\bibitem[Castillo et~al.(2015)Castillo, Schmidt-Hieber, and Van~der
  Vaart]{castillo2015bayesian}
Isma{\"e}l Castillo, Johannes Schmidt-Hieber, and Aad Van~der Vaart.
\newblock {Bayesian Linear Regression with Sparse Priors}.
\newblock \emph{The Annals of Statistics}, 43\penalty0 (5):\penalty0
  1986--2018, 2015.

\bibitem[Chandra et~al.(2020)Chandra, Canale, and Dunson]{chandra2020bayesian}
Noirrit~Kiran Chandra, Antonio Canale, and David~B Dunson.
\newblock {Bayesian Clustering of High-Dimensional Data}.
\newblock \emph{arXiv preprint arXiv:2006.02700}, 2020.

\bibitem[Dunson and Park(2008)]{dunson2008kernel}
David~B Dunson and Ju-Hyun Park.
\newblock {Kernel Stick-Breaking Processes}.
\newblock \emph{Biometrika}, 95\penalty0 (2):\penalty0 307--323, 2008.

\bibitem[Fraley and Raftery(2002)]{fraley2002model}
Chris Fraley and Adrian~E Raftery.
\newblock {Model-Based Clustering, Discriminant Analysis, and Density
  Estimation}.
\newblock \emph{Journal of the American Statistical Association}, 97\penalty0
  (458):\penalty0 611--631, 2002.

\bibitem[Gelman and Rubin(1992)]{gelman1992inference}
Andrew Gelman and Donald~B Rubin.
\newblock {Inference from Iterative Simulation Using Multiple Sequences}.
\newblock \emph{Statistical Science}, pages 457--472, 1992.

\bibitem[Heiner et~al.(2019)Heiner, Kottas, and Munch]{heiner2019structured}
Matthew Heiner, Athanasios Kottas, and Stephan Munch.
\newblock {Structured Priors for Sparse Probability Vectors with Application to
  Model Selection in Markov Chains}.
\newblock \emph{Statistics and Computing}, 29\penalty0 (5):\penalty0
  1077--1093, 2019.

\bibitem[Hoff(2009)]{hoff2009simulation}
Peter~D. Hoff.
\newblock {Simulation of the Matrix Bingham–von Mises–Fisher Distribution,
  with Applications to Multivariate and Relational Data}.
\newblock \emph{Journal of Computational and Graphical Statistics}, 18\penalty0
  (2):\penalty0 438--456, 2009.

\bibitem[Ishwaran and James(2001)]{ishwaran2001gibbs}
Hemant Ishwaran and Lancelot~F. James.
\newblock {Gibbs Sampling Methods for Stick-Breaking Priors}.
\newblock \emph{Journal of the American Statistical Association}, 96\penalty0
  (453):\penalty0 161--173, 2001.

\bibitem[Jain and Neal(2007)]{jain2007splitting}
Sonia Jain and Radford~M. Neal.
\newblock {Splitting and Merging Components of a Nonconjugate Dirichlet Process
  Mixture Model}.
\newblock \emph{Bayesian Analysis}, 2\penalty0 (3):\penalty0 445--472, 2007.

\bibitem[Kalli et~al.(2011)Kalli, Griffin, and Walker]{kalli2011slice}
Maria Kalli, Jim~E Griffin, and Stephen~G Walker.
\newblock {Slice Sampling Mixture Models}.
\newblock \emph{Statistics and computing}, 21\penalty0 (1):\penalty0 93--105,
  2011.

\bibitem[Marcus et~al.(2011)Marcus, Harwell, Olsen, Hodge, Glasser, Prior,
  Jenkinson, Laumann, Curtiss, and Van~Essen]{marcus2011informatics}
Daniel Marcus, John Harwell, Timothy Olsen, Michael Hodge, Matthew Glasser,
  Fred Prior, Mark Jenkinson, Timothy Laumann, Sandra Curtiss, and David
  Van~Essen.
\newblock {Informatics and Data Mining Tools and Strategies for the Human
  Connectome Project}.
\newblock \emph{Frontiers in neuroinformatics}, 5:\penalty0 4, 2011.

\bibitem[McLachlan et~al.(2003)McLachlan, Peel, and
  Bean]{mclachlan2003modelling}
Geoffrey~J McLachlan, David Peel, and Richard~W Bean.
\newblock {Modelling High-Dimensional Data by Mixtures of Factor Analyzers}.
\newblock \emph{Computational Statistics \& Data Analysis}, 41\penalty0
  (3-4):\penalty0 379--388, 2003.

\bibitem[Miller(2019)]{miller2019elementary}
Jeffrey~W Miller.
\newblock An elementary derivation of the chinese restaurant process from
  sethuraman’s stick-breaking process.
\newblock \emph{Statistics \& Probability Letters}, 146:\penalty0 112--117,
  2019.

\bibitem[Miller and Harrison(2013)]{miller2013simple}
Jeffrey~W Miller and Matthew~T Harrison.
\newblock {A Simple Example of Dirichlet Process Mixture Inconsistency for the
  Number of Components}.
\newblock In C.~J.~C. Burges, L.~Bottou, M.~Welling, Z.~Ghahramani, and K.~Q.
  Weinberger, editors, \emph{Advances in Neural Information Processing Systems
  26}, pages 199--206. Curran Associates, Inc., 2013.

\bibitem[Miller and Harrison(2014)]{miller2014inconsistency}
Jeffrey~W. Miller and Matthew~T. Harrison.
\newblock {Inconsistency of Pitman-Yor Process Mixtures for the Number of
  Components}.
\newblock \emph{Journal of Machine Learning Research}, 15\penalty0
  (96):\penalty0 3333--3370, 2014.

\bibitem[Miller and Harrison(2018)]{miller2018mixture}
Jeffrey~W. Miller and Matthew~T. Harrison.
\newblock {Mixture Models with a Prior on the Number of Components}.
\newblock \emph{Journal of the American Statistical Association}, 113\penalty0
  (521):\penalty0 340--356, 2018.

\bibitem[Nobile(1994)]{nobile1994bayesian}
Agostino Nobile.
\newblock \emph{{Bayesian Analysis of Finite Mixture Distributions}}.
\newblock PhD thesis, PhD Thesis. Carnegie Mellon University, Pittsburgh, 1994.

\bibitem[Ohn and Lin(2020)]{ohn2020optimal}
Ilsang Ohn and Lizhen Lin.
\newblock {Optimal Bayesian Estimation of Gaussian Mixtures with Growing Number
  of Components}.
\newblock \emph{arXiv preprint arXiv:2007.09284}, 2020.

\bibitem[Pitman(1995)]{pitman1995exchangeable}
Jim Pitman.
\newblock {Exchangeable and Partially Exchangeable Random Partitions}.
\newblock \emph{Probability Theory and Related Fields}, 102\penalty0
  (2):\penalty0 145--158, 1995.

\bibitem[Pitman and Yor(1997)]{pitman1997two}
Jim Pitman and Marc Yor.
\newblock {The Two-parameter Poisson-Dirichlet Distribution Derived from a
  Stable Subordinator}.
\newblock \emph{Annals of Probability}, 25\penalty0 (2):\penalty0 855--900,
  1997.

\bibitem[Ren et~al.(2011)Ren, Du, Carin, and Dunson]{ren2011logistic}
Lu~Ren, Lan Du, Lawrence Carin, and David~B Dunson.
\newblock {Logistic Stick-Breaking Process}.
\newblock \emph{Journal of Machine Learning Research}, 12\penalty0 (1), 2011.

\bibitem[Richardson and Green(1997)]{richardson1997bayesian}
Sylvia Richardson and Peter~J. Green.
\newblock {On Bayesian Analysis of Mixtures with an Unknown Number of
  Components (with discussion)}.
\newblock \emph{Journal of the Royal Statistical Society: Series B (Statistical
  Methodology)}, 59\penalty0 (4):\penalty0 731--792, 1997.

\bibitem[Rodr{\'\i}guez et~al.(2010)Rodr{\'\i}guez, Dunson, and
  Gelfand]{rodriguez2010latent}
Abel Rodr{\'\i}guez, David~B Dunson, and Alan~E Gelfand.
\newblock {Latent Stick-Breaking Processes}.
\newblock \emph{Journal of the American Statistical Association}, 105\penalty0
  (490):\penalty0 647--659, 2010.

\bibitem[Rohe et~al.(2011)Rohe, Chatterjee, and Yu]{rohe2011spectral}
Karl Rohe, Sourav Chatterjee, and Bin Yu.
\newblock {Spectral Clustering and the High-dimensional Stochastic Blockmodel}.
\newblock \emph{Annals of Statistics}, 39\penalty0 (4):\penalty0 1878--1915,
  2011.

\bibitem[Sethuraman(1994)]{sethuraman1994constructive}
Jayaram Sethuraman.
\newblock {A Constructive Definition of the Dirichlet Prior}.
\newblock \emph{Statistica Sinica}, 4:\penalty0 639--650, 1994.

\bibitem[Walker(2007)]{walker2007sampling}
Stephen~G Walker.
\newblock {Sampling the Dirichlet Mixture Model With Slices}.
\newblock \emph{Communications in Statistics —- Simulation and Computation},
  36\penalty0 (1):\penalty0 45--54, 2007.

\bibitem[Yang et~al.(2019)Yang, Xia, Ho, and Jordan]{yang2019posterior}
Chiao-Yu Yang, Eric Xia, Nhat Ho, and Michael~I Jordan.
\newblock {Posterior Distribution for the Number of Clusters in Dirichlet
  Process Mixture Models}.
\newblock \emph{arXiv preprint arXiv:1905.09959}, 2019.

\end{thebibliography}

\end{document}